\pdfoutput=1

\documentclass[conference,a4paper]{IEEEtran}
\usepackage[utf8]{inputenc} 
\usepackage[T1]{fontenc}
\usepackage{amsmath,amssymb,amsfonts}
\usepackage{algorithmic,algorithm}
\usepackage{tikz}
\usepackage{array}
\usepackage[caption=false,font=normalsize,labelfont=sf,textfont=sf]{subfig}
\usepackage{textcomp}
\usepackage{stfloats}
\usepackage{url}
\usepackage{verbatim}
\usepackage{graphicx}
\usepackage{cite}
\usepackage{arydshln} 
\usepackage{url}
\newtheorem{theorem}{\bf Theorem}
\newtheorem{lemma}[theorem]{\bf Lemma}
\newtheorem{corollary}[theorem]{\bf Corollary}
\newtheorem{definition}{\bf Definition}

\newtheorem{remark}{\bf Remark}
\begin{document}

\title{Two-Insertion/Deletion/Substitution Correcting Codes}

\author{
\IEEEauthorblockN{Yuhang Pi and Zhifang Zhang}
\IEEEauthorblockA{ 
Academy of Mathematics and Systems Science, Chinese Academy of Sciences, Beijing 100190, China\\ 
School of Mathematical Sciences, University of Chinese Academy of Sciences, Beijing 100049, China\\}
Email: piyuhang@amss.ac.cn, zfz@amss.ac.cn
}

\maketitle

\begin{abstract}
In recent years, the emergence of DNA storage systems has led to a widespread focus on the research of 
codes correcting insertions, deletions, and classic substitutions. During the initial investigation, 
Levenshtein discovered the VT codes are precisely capable of correcting single insertion/deletion 
and then extended the VT construction to single-insertion/deletion/substitution ($1$-ins/del/sub) 
correcting codes. Inspired by this, we generalize the recent findings of $1$-del $1$-sub 
correcting codes with redundancy $6\log_{2}n+O(1)$ to more general $2$-ins/del/sub correcting codes 
without increasing the redundancy. Our key technique is to apply higher-order VT syndromes to 
distinct objects and accomplish a systematic classification of all error patterns. 
\end{abstract}

\section{Introduction}\label{sec1}

The research of insertion/deletion/substitution (ins/del/sub) correcting codes is of 
significant interest not only in the field of communication but also in the field of life sciences, 
exhibiting a multitude of potential value in applications. Insertions, deletions, and substitutions 
commonly occur in DNA mutations \cite{Suz.2021}, as well as in the synthesizing and sequencing processes 
of a DNA storage system \cite{Hec.2019}. 

Levenshtein\cite{Lev.1966} discovered the binary Varshamov-Tenengolts (VT) codes \cite{Var.1965}, 
\begin{equation*}
VT_{a}(n)=\left\{ x_{1}\cdots x_{n}\in \{ 0,1 \}^{n}\mid 
\sum_{i=1}^{n}ix_{i}\equiv a \text{ mod } n+1 \right\},
\end{equation*}
are $1$-ins/del correcting codes with asymptotic optimal redundancy. 
If congruence modulo $n+1$ is replaced by congruence modulo $2n$, the corresponding codes 
become $1$-ins/del/sub correcting codes\cite{Lev.1966}. 
Furthermore, Levenshtein \cite{Lev.1967} constructed two-burst-deletion correcting codes which are also 
$1$-ins/del correcting codes, thereby signifying a complete extension from VT codes as well. 
Note that generally a two-burst-deletion correcting code may not necessarily be a $1$-ins/del 
correcting code (e.g., $\mathcal{C}=\{ 101,010 \}$). 

The Helberg codes proposed by Helberg and Ferreira \cite{Hel.2002} 
have been confirmed as multiple-ins/del correcting codes \cite{Abdel-Ghaffar.2012} with 
redundancy $\Omega(n)$ \cite{Pal.2012}. 
Based on hash function, Brakensiek \textit{et al.} \cite{Bra.2018} presented $t$-ins/del 
correcting codes with redundancy $O(t^{2}\log_{2}t\log_{2}n)$, 
followed by the work of Gabrys and Sala \cite{Gab.2019}, Sima and Bruck \cite{Sim.2021}, 
as well as Song \textit{et al.} \cite{Son.2022b}. 
However, considering that hash function heavily relies on exhaustive research, 
strictly speaking these codes do not possess explicit forms. 
In the subsequent discussion we will primarily concentrate on binary explicit-form codes related to 
$2$-ins/del/sub correcting codes. 

Sima \textit{et al.} \cite{Sim.2020} constructed $2$-ins/del correcting codes with redundancy 
$7\log_{2}n+o(\log_{2}n)$ from higher-order VT syndromes. It was improved by 
Guruswami and H\aa stad \cite{Gur.2021} to $2$-ins/del correcting codes with 
redundancy $4\log_{2}n+10\log_{2}(\log_{2}n)+O(1)$, 
which is currently the most superior construction. $2$-ins/del correcting codes 
with list size two and redundancy $3\log_{2}n+O(1)$ are also presented in \cite{Gur.2021}. 

Smagloy \textit{et al.} \cite{Sma.2023} constructed $1$-del $1$-sub correcting codes 
with redundancy $6\log_{2}n+O(1)$. The redundancy was reduced by a constant in \cite{Son.2022b}. 
$1$-del $1$-sub correcting codes with list size two were studied by 
Gabrys \textit{et al.} \cite{Gab.2023} and Song \textit{et al.} \cite{Son.2022a}. 

All of these explicit-form constructions adopted an extension technique of VT syndrome 
(i.e., higher-order VT syndromes) either directly or indirectly. Although this technique has been 
observed in the study of ins/del correcting codes \cite{Pal.2011}, its favorable properties 
were only recently confirmed by Sima \textit{et al.} \cite{Sim.2020,Sim.2021}, with a particular focus 
on the sign-preserving number of a target sequence (referring to Lemma \ref{Sim.l2.1}). 
The $1$-del $1$-sub correcting codes in \cite{Sma.2023} and the 
$1$-del multiple-sub correcting codes in \cite{Son.2022b} serve as the most representative examples, 
applying higher-order VT syndromes to each bit (or equivalently, the number of $1$s in an interval). 




\subsection{Our Contributions}\label{sec1.1}

We construct a family of $2$-ins/del/sub correcting codes (i.e., Theorem \ref{M.t3.2}), 
among which at least one has redundancy of at most $6\log_{2}n+8$ (i.e., Corollary \ref{M.c3.1}). 
Compared to previous work \cite{Sma.2023,Son.2022b}, we apply higher-order VT syndromes 
to the number of particular adjacent pairs in an interval. 

Under the requirement (i.e., Definition \ref{M.d2.1}) that the pairwise distances 
between all errors in two sequences are suitably large, we strictly define the types and type values 
of all errors and sequence pair (i.e., Definition \ref{M.d2.2} and Definition \ref{M.d2.3}). 
For the two sequences that may not satisfy the requirement, 
we employ a lemma (i.e., Lemma \ref{M.l2.1}) to analyze two other relevant sequences. 
In this way, we eliminate the confusion caused by errors being too close to each other and 
accomplish a rigorous classification. Based primarily on the type of sequence pair, 
we conduct a unified analysis of sign-preserving number to complete the proof. 
This analytical framework provides a novel perspective for the analysis of sign-preserving number, 
thereby facilitating the further applications of higher-order VT syndromes in code construction. 




\subsection{Organization}\label{sec1.2}

The rest of this paper is organized as follows. We introduce notations and known conclusions 
in Section \ref{sec2} and present main results in Section \ref{sec3}, 
including the construction of $2$-ins/del/sub correcting codes. 
In Sections \ref{sec4}--\ref{sec6}, we prove Theorem \ref{M.t3.1} by examining three cases, 
essentially verifying the code $\mathcal{C}_{k_{1},k_{2},k_{3},k_{4}}$ in Theorem \ref{M.t3.2} is 
a $2$-sub correcting code, a $2$-del correcting code, and a $1$-del $1$-sub correcting code, respectively. 
Section \ref{sec7} concludes this paper. 




\section{Preliminaries}\label{sec2}

Let $A_{2}=\{ 0,1 \}$, $[i]=\{ 1,2,\cdots,i \}$, $[i,j]=\{ i,i+1,\cdots,j \}$, and 
$\mathbf{VT}_{i}^{n}=(1^{i},2^{i},\cdots,n^{i})$. For $\mathbf{x}\in A_{2}^{n}$, 
denoted by $x_{i}$ the $i$-th symbol in $\mathbf{x}$, by $\mathcal{B}_{t,s,r}(\mathbf{x})$ 
the set of sequences which can be obtained from $\mathbf{x}$ by 
$t$ insertions, $s$ deletions, and $r$ substitutions. Trivial substitution at $x_{i}$, 
i.e., no change occurring at $x_{i}$, can also be regarded as one substitution. 
Sequence $x_{1}\cdots x_{n}$ and vector $(x_{1},\cdots,x_{n})$ are commonly regarded as consistent. 
If $S=\{ i_{1},\cdots,i_{j} \}\subseteq [n]$ where $i_{1}< i_{2}<\cdots <i_{j}$, 
$\mathbf{x}_{S}$ denotes $x_{i_{1}}\cdots x_{i_{j}}$. 
Edit distance $L^{\ast}(\mathbf{x},\mathbf{y})$ of $\mathbf{x}$ and $\mathbf{y}$ is defined as the 
minimum number of insertions, deletions, and substitutions needed to transform $\mathbf{x}$ into $\mathbf{y}$. 
$\mathcal{C}\subseteq A_{2}^{n}$ is called a $k$-ins/del/sub correcting code if 
$L^{\ast}(\mathbf{c_{1}},\mathbf{c_{2}})\geq 2k+1$ for all 
$\mathbf{c_{1}},\mathbf{c_{2}}\in\mathcal{C}, \mathbf{c_{1}}\neq \mathbf{c_{2}}$. 

Use $a$ to represent any one of $0$ and $1$, $b$ to represent another one. 
For $\mathbf{x}\in A_{2}^{n}$, 
$f(\mathbf{x})$ is defined as the number of adjacent pair $ab$ in $\mathbf{x}$, 
i.e., the total number of adjacent pairs $01$ and $10$ in $\mathbf{x}$. 
Furthermore, let $\mathbf{F}(\mathbf{x})=(f(\mathbf{x}_{[1]}),\cdots,f(\mathbf{x}_{[n]}))$. 
For instance, $\mathbf{F}(000100)=(0,0,0,1,2,2)$. 

If a lowercase bold letter represents a sequence, 
the sequence with a $0$ added at both the beginning and the end of it, is denoted by 
the corresponding uppercase bold letter. For instance, $\mathbf{x}=110$ corresponds to $\mathbf{X}=01100$. 
Similar techniques are applied to both \cite{Lev.1967} and \cite{Gur.2021}. 
The advantages of adding two $0$s are twofold. Firstly, it reduces the number of cases to be discussed later. 
Secondly, it ensures our capability to distinguish $\mathbf{x}$ and $\mathbf{y}$. 
Specifically, if $\mathbf{x}\neq \mathbf{y}$, 
it is possible that $\mathbf{F}(\mathbf{x})=\mathbf{F}(\mathbf{y})$ 
(e.g., $\mathbf{x}=10$ and $\mathbf{y}=01$). However, $\mathbf{x}\neq \mathbf{y}$ is equivalent to 
$\mathbf{F}(\mathbf{X})\neq \mathbf{F}(\mathbf{Y})$. 
This conclusion can be reached by focusing on the first unequal symbols of $\mathbf{X}$ and $\mathbf{Y}$. 

If an integer sequence $\mathbf{z}=z_{1}\cdots z_{n}$ is 
non-negative or non-positive, $\mathbf{z}$ is called a $1$-sequence. 
If an integer sequence $\mathbf{z}$ can be divided into $k$ continuous segments such that each segment 
is a $1$-sequence, $\mathbf{z}$ is called a $k$-sequence. 
The sign-preserving number of $\mathbf{z}$, denoted by $\sigma(\mathbf{z})$, 
is defined as the minimum integer $k$ such that $\mathbf{z}$ is a $k$-sequence. 
For instance, $\sigma((1,0,1,-1,-2,3))=3$. Clearly, for $1\leq i\leq n-1$, 
$\sigma(z_{1}\cdots z_{n})\leq \sigma(z_{1}\cdots z_{i})+\sigma(z_{i+1}\cdots z_{n})$. 
Introducing the concept of sign-preserving number has the benefit of characterizing the following lemma. 

\begin{lemma}[c.f., \cite{Sim.2020,Sim.2021}]\label{Sim.l2.1}
If $\mathbf{z}\in\mathbb{Z}^{n}$ satisfies $\mathbf{z}\cdot \mathbf{VT}_{i}^{n}=0$ 
for $0\leq i\leq \sigma(\mathbf{z})-1$, then $\mathbf{z}=0^{n}$. 
\end{lemma}

\begin{definition}\label{M.d2.1}
Let $\mathbf{u},\mathbf{v}\in A_{2}^{m}$. 
$(\mathbf{U},\mathbf{V})$ is called a $(s,r)$-del/sub good pair if 
there exist $i_{1},i_{2},\cdots,i_{2s+2r}\in [2,m+1]$ with pairwise distances 
at least $2s+1$ such that the sequence obtained 
by deleting $U_{i_{1}},\cdots,U_{i_{s}}$ and substituting $U_{i_{s+1}},\cdots,U_{i_{s+2r}}$ 
(allowing trivial substitutions) from $\mathbf{U}$ is equal to the sequence obtained by 
deleting $V_{i_{s+2r+1}},\cdots,V_{i_{2s+2r}}$ from $\mathbf{V}$. 
\end{definition}

\begin{remark}
Levenshtein \cite{Lev.1966} proved the equivalence between $t$-ins correcting codes, 
$t$-del correcting codes, and $t$-ins/del correcting codes. 
Similarly, the equivalence between $t$-ins $s$-sub correcting codes, 
$t$-del $s$-sub correcting codes, and $t$-ins/del $s$-sub correcting codes still holds. 
Therefore, we only need to consider deletions and substitutions in Definition \ref{M.d2.1}. 
Moreover, for $(\mathbf{X},\mathbf{Y})$ that may not be a $(s,r)$-del/sub good pair, 
we will later shift our analysis to a relevant $(s,r)$-del/sub good pair $(\mathbf{U},\mathbf{Y})$ 
through Lemma \ref{M.l2.1}. 
\end{remark}

Assume $(\mathbf{U},\mathbf{V})$ is a $(s,r)$-del/sub good pair. 
Consequently, for any $\alpha\in [m+2]\backslash\{  i_{1},\cdots,i_{s} \}$, 
there exists a unique $\beta\in [m+2]\backslash \{ i_{s+2r+1},\cdots,i_{2s+2r} \}$ 
such that $U_{\alpha}$ and $V_{\beta}$ are matched. 
we express $\tau(U_{\alpha})$ as $V_{\beta}$, and conversely, 
$\tau(V_{\beta})$ as $U_{\alpha}$. 
For $\alpha\in\{ i_{1},\cdots,i_{s} \}$ and $\beta\in\{ i_{s+2r+1},\cdots,i_{2s+2r} \}$, 
$\tau(U_{\alpha})$ and $\tau(V_{\beta})$ are undefined. 
This notation is informal, but highly flexible. 

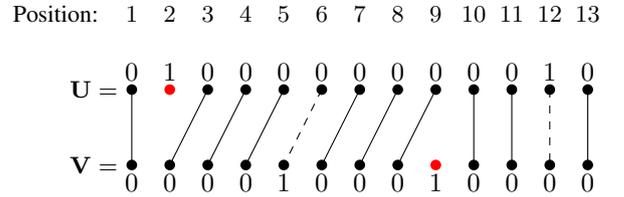
\begin{figure}[htbp]
\centering
\begin{tikzpicture}

\node at (-0.5,1.5) {\small Position:};

\node at (0.5,1.5) {\small $1$};

\node at (1,1.5) {\small $2$};

\node at (1.5,1.5) {\small $3$};

\node at (2,1.5) {\small $4$};

\node at (2.5,1.5) {\small $5$};

\node at (3,1.5) {\small $6$};

\node at (3.5,1.5) {\small $7$};

\node at (4,1.5) {\small $8$};

\node at (4.5,1.5) {\small $9$};

\node at (5,1.5) {\small $10$};

\node at (5.5,1.5) {\small $11$};

\node at (6,1.5) {\small $12$};

\node at (6.5,1.5) {\small $13$};

\node at (0,0.5) {\small$\mathbf{U}=$};

\fill (0.5,0.5) circle (2pt);
\node [above] at (0.5,0.5) {$0$};

\fill[red] (1,0.5) circle (2pt);
\node [above] at (1,0.5) {$1$};

\fill (1.5,0.5) circle (2pt);
\node [above] at (1.5,0.5) {$0$};

\fill (2,0.5) circle (2pt);
\node [above] at (2,0.5) {$0$};

\fill (2.5,0.5) circle (2pt);
\node [above] at (2.5,0.5) {$0$};

\fill (3,0.5) circle (2pt);
\node [above] at (3,0.5) {$0$};

\fill (3.5,0.5) circle (2pt);
\node [above] at (3.5,0.5) {$0$};

\fill (4,0.5) circle (2pt);
\node [above] at (4,0.5) {$0$};

\fill (4.5,0.5) circle (2pt);
\node [above] at (4.5,0.5) {$0$};

\fill (5,0.5) circle (2pt);
\node [above] at (5,0.5) {$0$};

\fill (5.5,0.5) circle (2pt);
\node [above] at (5.5,0.5) {$0$};

\fill (6,0.5) circle (2pt);
\node [above] at (6,0.5) {$1$};

\fill (6.5,0.5) circle (2pt);
\node [above] at (6.5,0.5) {$0$};

\node at (0,-0.5) {\small$\mathbf{V}=$};

\fill (0.5,-0.5) circle (2pt);
\node [below] at (0.5,-0.5) {$0$};

\fill (1,-0.5) circle (2pt);
\node [below] at (1,-0.5) {$0$};

\fill (1.5,-0.5) circle (2pt);
\node [below] at (1.5,-0.5) {$0$};

\fill (2,-0.5) circle (2pt);
\node [below] at (2,-0.5) {$0$};

\fill (2.5,-0.5) circle (2pt);
\node [below] at (2.5,-0.5) {$1$};

\fill (3,-0.5) circle (2pt);
\node [below] at (3,-0.5) {$0$};

\fill (3.5,-0.5) circle (2pt);
\node [below] at (3.5,-0.5) {$0$};

\fill (4,-0.5) circle (2pt);
\node [below] at (4,-0.5) {$0$};

\fill[red] (4.5,-0.5) circle (2pt);
\node [below] at (4.5,-0.5) {$1$};

\fill (5,-0.5) circle (2pt);
\node [below] at (5,-0.5) {$0$};

\fill (5.5,-0.5) circle (2pt);
\node [below] at (5.5,-0.5) {$0$};

\fill (6,-0.5) circle (2pt);
\node [below] at (6,-0.5) {$0$};

\fill (6.5,-0.5) circle (2pt);
\node [below] at (6.5,-0.5) {$0$};

\draw (0.5,0.5)--(0.5,-0.5);

\draw (1.5,0.5)--(1,-0.5);

\draw (2,0.5)--(1.5,-0.5);

\draw (2.5,0.5)--(2,-0.5);

\draw[dashed] (3,0.5)--(2.5,-0.5);

\draw (3.5,0.5)--(3,-0.5);

\draw (4,0.5)--(3.5,-0.5);

\draw (4.5,0.5)--(4,-0.5);

\draw (5,0.5)--(5,-0.5);

\draw (5.5,0.5)--(5.5,-0.5);

\draw[dashed] (6,0.5)--(6,-0.5);

\draw (6.5,0.5)--(6.5,-0.5);

\end{tikzpicture}
\caption{The matching of $\mathbf{U}_{[13]\backslash\{ 2 \}}$ and $\mathbf{V}_{[13]\backslash\{ 9 \}}$.}
\label{M.f2.1}
\end{figure}

\begin{figure}[htbp]
\centering
\begin{tikzpicture}

\node at (-0.5,1.5) {\small Position:};

\node at (0.5,1.5) {\small $1$};

\node at (1,1.5) {\small $2$};

\node at (1.5,1.5) {\small $3$};

\node at (2,1.5) {\small $4$};

\node at (2.5,1.5) {\small $5$};

\node at (3,1.5) {\small $6$};

\node at (3.5,1.5) {\small $7$};

\node at (4,1.5) {\small $8$};

\node at (0,0.5) {\small$\mathbf{U}=$};

\fill (0.5,0.5) circle (2pt);
\node [above] at (0.5,0.5) {$0$};

\fill (1,0.5) circle (2pt);
\node [above] at (1,0.5) {$0$};

\fill (1.5,0.5) circle (2pt);
\node [above] at (1.5,0.5) {$0$};

\fill (2,0.5) circle (2pt);
\node [above] at (2,0.5) {$0$};

\fill (2.5,0.5) circle (2pt);
\node [above] at (2.5,0.5) {$0$};

\fill (3,0.5) circle (2pt);
\node [above] at (3,0.5) {$1$};

\fill (3.5,0.5) circle (2pt);
\node [above] at (3.5,0.5) {$1$};

\fill (4,0.5) circle (2pt);
\node [above] at (4,0.5) {$0$};

\node at (0,-0.5) {\small$\mathbf{V}=$};

\fill (0.5,-0.5) circle (2pt);
\node [below] at (0.5,-0.5) {$0$};

\fill (1,-0.5) circle (2pt);
\node [below] at (1,-0.5) {$1$};

\fill (1.5,-0.5) circle (2pt);
\node [below] at (1.5,-0.5) {$1$};

\fill (2,-0.5) circle (2pt);
\node [below] at (2,-0.5) {$0$};

\fill (2.5,-0.5) circle (2pt);
\node [below] at (2.5,-0.5) {$0$};

\fill (3,-0.5) circle (2pt);
\node [below] at (3,-0.5) {$0$};

\fill (3.5,-0.5) circle (2pt);
\node [below] at (3.5,-0.5) {$0$};

\fill (4,-0.5) circle (2pt);
\node [below] at (4,-0.5) {$0$};

\draw (0.5,0.5)--(0.5,-0.5);

\draw[dashed] (1,0.5)--(1,-0.5);

\draw[dashed] (1.5,0.5)--(1.5,-0.5);

\draw (2,0.5)--(2,-0.5);

\draw (2.5,0.5)--(2.5,-0.5);

\draw[dashed] (3,0.5)--(3,-0.5);

\draw[dashed] (3.5,0.5)--(3.5,-0.5);

\draw (4,0.5)--(4,-0.5);

\end{tikzpicture}
\caption{The matching of $\mathbf{U}$ and $\mathbf{V}$.}
\label{M.f2.2}
\end{figure}
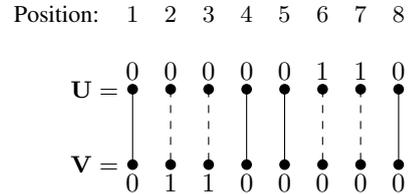

As an example shown in Fig. \ref{M.f2.1}, the deleted symbols are marked with red dots, 
while the lines indicate the one-to-one matching between $\mathbf{U}_{[13]\backslash\{ 2 \}}$ and 
$\mathbf{V}_{[13]\backslash\{ 9 \}}$. Particularly, two dashed lines indicate two substitutions. 
At this moment, $\tau(U_{3})=V_{2}=0$ and $\tau(U_{6})=V_{5}=1$. However, 
$\tau(U_{2})$ and $\tau(V_{9})$ are undefined. 
As another example shown in Fig. \ref{M.f2.2}, $\tau(U_{i})=V_{i}$ and $\tau(V_{i})=U_{i}$ for $i\in [8]$. 

For $(s,r)$-del/sub good pair $(\mathbf{U},\mathbf{V})$, 
all errors can be regarded as relatively independent parts, 
enabling their classification into types and type values in Definition \ref{M.d2.2}. 
Additionally, there exists an evident sequential order among all errors, which enables us 
to define the type and type value of $(\mathbf{U},\mathbf{V})$ in Definition \ref{M.d2.3}. 

\begin{definition}\label{M.d2.2}
Assume $(\mathbf{U},\mathbf{V})$ is a $(s,r)$-del/sub good pair introduced in Definition \ref{M.d2.1}. 
The types and type values of all errors are defined as follows. 
\begin{enumerate}
\item 
The type and type value of substitution occurring at $U_{i}$ where $i\in\{ i_{s+1},\cdots,i_{s+2r} \}$ 
is defined as $sub$ and $e$ respectively where 
$e=f(\tau(U_{i-1})U_{i}U_{i+1})-f(\tau(U_{i-1})\tau(U_{i})U_{i+1})$. 

\item
The type and type value of deletion occurring at $U_{i}$ where $i\in\{ i_{1},\cdots,i_{s} \}$ 
is defined as $\overline{del}$ and $\overline{e}$ respectively where 
$e=f(U_{i-1}U_{i}U_{i+1})-f(\tau(U_{i-1})\tau(U_{i+1}))
=f(U_{i-1}U_{i}U_{i+1})-f(U_{i-1}U_{i+1})$. 

\item
The type and type value of deletion occurring at $V_{i}$ where $i\in\{ i_{s+2r+1},\cdots,i_{2s+2r} \}$ 
is defined as $\underline{del}$ and $\underline{e}$ respectively where 
$e=f(\tau(V_{i-1})\tau(V_{i+1}))-f(V_{i-1}V_{i}V_{i+1})
=f(V_{i-1}V_{i+1})-f(V_{i-1}V_{i}V_{i+1})$. 
\end{enumerate}
\end{definition}

Due to the separation of errors, the definitions and properties of 
$\tau(U_{i-1})$, $\tau(U_{i+1})$, $\tau(V_{i-1})$, and $\tau(V_{i+1})$ 
in Definition \ref{M.d2.2} are valid. 

\begin{table}[htbp]
\caption{The Types and Type Values of Errors. 
(A) Substitutions in $\mathbf{U}$. (B) Deletions in $\mathbf{U}$. (C) Deletions in $\mathbf{V}$.}
\label{table1}
\begin{center}
\begin{tabular}{|c|c|c|}
\hline
$\tau(U_{i-1})U_{i}U_{i+1}\rightarrow\tau(U_{i-1})\tau(U_{i})U_{i+1}$ & Types & Type Values
\\
\hline
Trivial Substitution & $sub$ & $0$ 
\\
\hline
$a\mathbf{b}b \rightarrow a\mathbf{a}b$ & $sub$ & $0$ 
\\
\hline
$a\mathbf{a}b \rightarrow a\mathbf{b}b$ & $sub$ & $0$ 
\\
\hline
$a\mathbf{b}a \rightarrow a\mathbf{a}a$ & $sub$ & $2$ 
\\
\hline
$a\mathbf{a}a \rightarrow a\mathbf{b}a$ & $sub$ & $-2$ 
\\
\hline
\end{tabular}
\vspace{0.2cm}

(a)

\vspace{0.5cm}

\begin{tabular}{|c|c|c|}
\hline
$U_{i-1}U_{i}U_{i+1}\rightarrow U_{i-1}U_{i+1}$ & Types & Type Values
\\
\hline
$a\mathbf{b}b \rightarrow ab$ & \raisebox{-1.7pt}{$\overline{del}$} & \raisebox{-1.5pt}{$\overline{0}$} 
\\
\hline
$a\mathbf{a}b \rightarrow ab$ & \raisebox{-1.7pt}{$\overline{del}$} & \raisebox{-1.5pt}{$\overline{0}$} 
\\
\hline
$a\mathbf{a}a \rightarrow aa$ & \raisebox{-1.7pt}{$\overline{del}$} & \raisebox{-1.5pt}{$\overline{0}$} 
\\
\hline
$a\mathbf{b}a \rightarrow aa$ & \raisebox{-1.7pt}{$\overline{del}$} & \raisebox{-1.5pt}{$\overline{2}$} 
\\
\hline
\end{tabular}
\vspace{0.2cm}

(b)

\vspace{0.5cm}

\begin{tabular}{|c|c|c|}
\hline
$V_{i-1}V_{i}V_{i+1}\rightarrow V_{i-1}V_{i+1}$ & Types & Type Values
\\
\hline
$a\mathbf{b}b \rightarrow ab$ & $\underline{del}$ & $\underline{0}$ 
\\
\hline
$a\mathbf{a}b \rightarrow ab$ & $\underline{del}$ & $\underline{0}$ 
\\
\hline
$a\mathbf{a}a \rightarrow aa$ & $\underline{del}$ & $\underline{0}$ 
\\
\hline
$a\mathbf{b}a \rightarrow aa$ & $\underline{del}$ & $\underline{-2}$ 
\\
\hline
\end{tabular}
\vspace{0.2cm}

(c)

\end{center}
\end{table}

Table \ref{table1} shows all types and type values of substitutions and deletions, 
with the bold symbols indicating the occurrences of corresponding errors. 
Although a substitution of type value $0$ occurring at $U_{i}$ 
does not affect the number of adjacent pair $ab$ of the entire sequence, 
it may alter the number of adjacent pair $ab$ of the first $i$ symbols, 
with a difference of at most $1$. 

\begin{definition}\label{M.d2.3}
Assume $(\mathbf{U},\mathbf{V})$ is a $(s,r)$-del/sub good pair introduced in Definition \ref{M.d2.1}. 
The type of $(\mathbf{U},\mathbf{V})$ belongs to $\{ sub,\overline{del},\underline{del} \}^{2s+2r}$, 
determined in sequential order according to the types of all errors. 
The type value of $(\mathbf{U},\mathbf{V})$ belongs to 
$\{ 2,0,-2,\overline{2},\overline{0},\underline{0},\underline{-2} \}^{2s+2r}$, 
determined in sequential order according to the type values of all errors. 
\end{definition}

Fig. \ref{M.f2.1} corresponds to $s=1$, $r=1$, $i_{1}=2$, $i_{2}=6$, $i_{3}=12$, and $i_{4}=9$. 
Given that the pairwise distances between $i_{1},\cdots,i_{4}$ are at least $2s+1$, 
$(\mathbf{U},\mathbf{V})$ is a $(1,1)$-del/sub good pair. 
Therefore, the types and type values of errors can be determined. 
The deletion occurring at $U_{2}$ is of type $\overline{del}$ and type value $\overline{2}$. 
The substitution occurring at $U_{6}$ is of type $sub$ and type value $-2$. 
The deletion occurring at $V_{9}$ is of type $\underline{del}$ and type value $\underline{-2}$. 
The substitution occurring at $U_{12}$ is of type $sub$ and type value $2$. 
Based on them, we ascertain that the type and type value of $(\mathbf{U},\mathbf{V})$ 
are $(\overline{del},sub,\underline{del},sub)$ and $(\overline{2},-2,\underline{-2},2)$, respectively. 

Fig. \ref{M.f2.2} corresponds to $s=0$, $r=2$, $i_{1}=2$, $i_{2}=3$, $i_{3}=6$, and $i_{4}=7$. 
Given that the pairwise distances between $i_{1},\cdots,i_{4}$ are at least $2s+1$, 
$(\mathbf{U},\mathbf{V})$ is a $(0,2)$-del/sub good pair. 
Therefore, the types and type values of errors can be determined. 
The substitution occurring at $U_{2}$ is of type $sub$ and type value $-2$. 
The substitution occurring at $U_{3}$ is of type $sub$ and type value $0$. 
The substitution occurring at $U_{6}$ is of type $sub$ and type value $0$. 
The substitution occurring at $U_{7}$ is of type $sub$ and type value $2$. 
Based on them, we ascertain that the type and type value of $(\mathbf{U},\mathbf{V})$ 
are $(sub,sub,sub,sub)$ and $(-2,0,0,2)$, respectively. 

Next, we revert to general $\mathbf{x},\mathbf{y}\in A_{2}^{n}$, 
$\mathcal{B}_{0,s,r}(\mathbf{x})\cap \mathcal{B}_{0,s,r}(\mathbf{y})\neq\varnothing$, 
where $n\geq 7$ and $s+r=2$. 
In the case of $s=0$, due to the permission of trivial substitutions, 
$(\mathbf{X},\mathbf{Y})$ is a $(0,2)$-del/sub good pair. 
In the case of $1\leq s\leq 2$, we employ the following lemma to analyze $(\mathbf{U},\mathbf{V})$ 
instead of $(\mathbf{X},\mathbf{Y})$, which satisfies the requirement that 
all errors are separated with pairwise distances at least $2s+1$. 

\begin{lemma}\label{M.l2.1}
For $\mathbf{x},\mathbf{y}\in A_{2}^{n}$, 
$\mathcal{B}_{0,s,r}(\mathbf{x})\cap \mathcal{B}_{0,s,r}(\mathbf{y})\neq \varnothing$, 
and any positive integer $k$, there exist $\mathbf{u},\mathbf{v}\in A_{2}^{m}$, 
such that $f(\mathbf{X})-f(\mathbf{Y})=f(\mathbf{U})-f(\mathbf{V})$ and 
$\sigma(\mathbf{F}(\mathbf{X})-\mathbf{F}(\mathbf{Y}))\leq 
\sigma(\mathbf{F}(\mathbf{U})-\mathbf{F}(\mathbf{V}))$. 
Moreover, there exist $i_{1},i_{2},\cdots,i_{2s+2r}\in [2,m+1]$ with pairwise distances at least $k$, 
such that the sequence obtained by deleting $U_{i_{1}},\cdots,U_{i_{s}}$ and substituting 
$U_{i_{s+1}},\cdots,U_{i_{s+2r}}$ (allowing trivial substitutions) from $\mathbf{U}$ is equal to 
the sequence obtained by deleting $V_{i_{s+2r+1}},\cdots,V_{i_{2s+2r}}$ from $\mathbf{V}$. 
\end{lemma}

The proof of Lemma \ref{M.l2.1} is presented in Appendix. 




\section{Main Results}\label{sec3}

\begin{theorem}\label{M.t3.1}
Let $n\geq 7$. If $\mathbf{x},\mathbf{y}\in A_{2}^{n}$ satisfy 
$L^{\ast}(\mathbf{x},\mathbf{y})\leq 4$ and 
\begin{equation}\label{M.e3.1}
\begin{cases}
(\mathbf{F}(\mathbf{X})-\mathbf{F}(\mathbf{Y}))\cdot \mathbf{VT}_{0}^{n+2}=0
\\
(\mathbf{F}(\mathbf{X})-\mathbf{F}(\mathbf{Y}))\cdot \mathbf{VT}_{1}^{n+2}=0
\\
(\mathbf{F}(\mathbf{X})-\mathbf{F}(\mathbf{Y}))\cdot \mathbf{VT}_{2}^{n+2}=0
\\
f(\mathbf{X})=f(\mathbf{Y})
\end{cases},
\end{equation}
then $\mathbf{x}=\mathbf{y}$.
\end{theorem}

The proof of Theorem \ref{M.t3.1} will be systematically developed in Sections \ref{sec4}--\ref{sec6}. 
Note that under the conditions in \eqref{M.e3.1}, 
$\sigma(\mathbf{F}(\mathbf{X})-\mathbf{F}(\mathbf{Y}))\leq 3$ is a sufficient condition for 
$\mathbf{x}=\mathbf{y}$ by Lemma \ref{Sim.l2.1}. 
In the rest of this section, under the premise of Theorem \ref{M.t3.1}, 
we present a family of $2$-ins/del/sub correcting codes. 
\begin{theorem}\label{M.t3.2}
For $n\geq 7$, the binary code $\mathcal{C}_{k_{1},k_{2},k_{3},k_{4}}$ in which the codeword 
$\mathbf{x}\in A_{2}^{n}$ satisfies 
\begin{equation}\label{M.e3.2}
\begin{cases}
\mathbf{F}(\mathbf{X})\cdot \mathbf{VT}_{0}^{n+2}\equiv k_{1} \text{ mod } 4n
\\
\mathbf{F}(\mathbf{X})\cdot \mathbf{VT}_{1}^{n+2}\equiv k_{2} \text{ mod } 2n^{2}
\\
\mathbf{F}(\mathbf{X})\cdot \mathbf{VT}_{2}^{n+2}\equiv k_{3} \text{ mod } 2n^{3}
\\
f(\mathbf{X})\equiv k_{4} \text{ mod } 9
\end{cases}
\end{equation}
is a $2$-ins/del/sub correcting code.
\end{theorem}

\begin{IEEEproof}
Suppose there exist $\mathbf{x}\neq\mathbf{y}\in \mathcal{C}_{k_{1},k_{2},k_{3},k_{4}}$ such that 
$L^{\ast}(\mathbf{x},\mathbf{y})\leq 4$. We only need to derive a contradiction. 
For the sake of simplicity, in the rest of this proof, 
$\mathbf{F}$ and $f_{i}$ denote $\mathbf{F}(\mathbf{X})-\mathbf{F}(\mathbf{Y})$ 
and $f(\mathbf{X}_{[i]})-f(\mathbf{Y}_{[i]})$, respectively. 

By \eqref{M.e3.2}, 
\begin{equation}\label{M.e3.3}
\begin{cases}
\mathbf{F}\cdot \mathbf{VT}_{0}^{n+2}\equiv 0 \text{ mod } 4n
\\
\mathbf{F}\cdot \mathbf{VT}_{1}^{n+2}\equiv 0 \text{ mod } 2n^{2}
\\
\mathbf{F}\cdot \mathbf{VT}_{2}^{n+2}\equiv 0 \text{ mod } 2n^{3}
\\
f_{n+2}\equiv 0 \text{ mod } 9
\end{cases}.
\end{equation}

By definition, 
$f_{1}=f(\mathbf{X}_{[1]})-f(\mathbf{Y}_{[1]})=0-0=0$. 
Moreover, an insertion, deletion, or substitution occurring at $\mathbf{X}$ will change $f(\mathbf{X})$ 
by at most $2$, which implies $|f_{n+2}|\leq 8$ and thereby $f_{n+2}=0$. 
Hence, $|f_{2}|,|f_{n+1}|\leq 1$, $|f_{3}|,|f_{n}|\leq 2$, and $|f_{4}|,|f_{n-1}|\leq 3$. 
Due to $f_{n+2}=0$, $|f_{i}|\leq 4$ holds for $5\leq i\leq n-2$. Thus, 
\begin{equation}\label{M.i3.1}
\begin{split}
|\mathbf{F}\cdot \mathbf{VT}_{i}^{n+2}|
\leq&\sum_{j=1}^{n+2}|f_{j}|\cdot j^{i}
\\
\leq&1\cdot 2^{i}+2\cdot 3^{i}+3\cdot 4^{i}+4\cdot\sum_{j=5}^{n-2}j^{i}
\\
&+3\cdot (n-1)^{i}
+2\cdot n^{i}+1\cdot (n+1)^{i}.
\end{split}
\end{equation}

Replacing $i$ with $0,1,2$ in \eqref{M.i3.1}, we obtain 
\begin{equation}\label{M.i3.2}
\begin{cases}
|\mathbf{F}\cdot \mathbf{VT}_{0}^{n+2}|\leq 4n-12<4n
\\
|\mathbf{F}\cdot \mathbf{VT}_{1}^{n+2}|\leq 2n^{2}-18<2n^{2}
\\
|\mathbf{F}\cdot \mathbf{VT}_{2}^{n+2}|\leq \frac{4n^{3}}{3}+\frac{14n}{3}-50<2n^{3}
\end{cases}.
\end{equation}

Combining \eqref{M.e3.3}, \eqref{M.i3.2}, and Theorem \ref{M.t3.1}, $\mathbf{x}=\mathbf{y}$ holds, 
a contradiction. 
\end{IEEEproof}

\begin{corollary}\label{M.c3.1}
For $n\geq 7$, there exist appropriate integers $k_{1},k_{2},k_{3},k_{4}$ such that the code 
$\mathcal{C}_{k_{1},k_{2},k_{3},k_{4}}$ presented in Theorem \ref{M.t3.2} is a 
$2$-ins/del/sub correcting code with redundancy of at most $6\log_{2}n+8$.
\end{corollary}

\begin{IEEEproof}
By Theorem \ref{M.t3.2} and the pigeonhole principle.
\end{IEEEproof}





\section{$\mathcal{C}_{k_{1},k_{2},k_{3},k_{4}}$ is a $2$-Sub Correcting Code}\label{sec4}

In this section, we assume the conditions in Theorem \ref{M.t3.1} hold, 
and proceed to verify $\mathbf{x}=\mathbf{y}$ in the case of 
$\mathcal{B}_{0,0,2}(\mathbf{x})\cap \mathcal{B}_{0,0,2}(\mathbf{y})\neq \varnothing$. 
Specifically, $\mathbf{Y}$ can be obtained from 
$\mathbf{X}$ by substituting $X_{j_{1}}$, $X_{j_{2}}$, $X_{j_{3}}$, and $X_{j_{4}}$ where 
$2\leq j_{1}<j_{2}<j_{3}<j_{4}\leq n+1$ (allowing trivial substitutions). 
For the sake of simplicity, in the rest of this section, 
$\mathbf{F}$ and $f_{i}$ denote $\mathbf{F}(\mathbf{X})-\mathbf{F}(\mathbf{Y})$ 
and $f(\mathbf{X}_{[i]})-f(\mathbf{Y}_{[i]})$, respectively. 

As a result of $f_{n+2}=0$, the number of substitutions of type value $2$ is equal to 
the number of substitutions of type value $-2$. We divide our discussion into three subcases. 




\subsection{Four Substitutions of Type Value $0$}\label{sec4.1}

In this situation, the conditions specified at the beginning of this section are transformed into 
\begin{equation}\label{M.e4.1}
\begin{cases}
2\leq j_{1}<j_{2}<j_{3}<j_{4}\leq n+1
\\
f_{j_{1}},f_{j_{2}},f_{j_{3}},f_{j_{4}}\in\{ -1,0,1 \}
\\
f_{j_{1}}+f_{j_{2}}+f_{j_{3}}+f_{j_{4}}=0
\\
j_{1}f_{j_{1}}+j_{2}f_{j_{2}}+j_{3}f_{j_{3}}+j_{4}f_{j_{4}}=0
\\
j_{1}^{2}f_{j_{1}}+j_{2}^{2}f_{j_{2}}+j_{3}^{2}f_{j_{3}}+j_{4}^{2}f_{j_{4}}=0
\end{cases}.
\end{equation}

\begin{lemma}\label{M.l4.1}
The equation in \eqref{M.e4.1} only has one solution $f_{j_{1}}=f_{j_{2}}=f_{j_{3}}=f_{j_{4}}=0$. 
\end{lemma}

\begin{IEEEproof}
Assume at least one of $f_{j_{1}}$, $f_{j_{2}}$, $f_{j_{3}}$, and $f_{j_{4}}$ is non-zero. 
Then the first four conditions in \eqref{M.e4.1} result in 
\begin{equation*}
\begin{cases}
f_{j_{1}}=f_{j_{4}}=1
\\
f_{j_{2}}=f_{j_{3}}=-1
\\
j_{1}+j_{4}=j_{2}+j_{3}
\end{cases},
\end{equation*}
or 
\begin{equation*}
\begin{cases}
f_{j_{1}}=f_{j_{4}}=-1
\\
f_{j_{2}}=f_{j_{3}}=1
\\
j_{1}+j_{4}=j_{2}+j_{3}
\end{cases}.
\end{equation*}

In all cases, the fifth condition in \eqref{M.e4.1} changes to 
$j_{1}^{2}+j_{4}^{2}=j_{2}^{2}+j_{3}^{2}$. Additionally, 
\begin{equation*}
j_{1}j_{4}
=\frac{(j_{1}+j_{4})^{2}-(j_{1}^{2}+j_{4}^{2})}{2}
=\frac{(j_{2}+j_{3})^{2}-(j_{2}^{2}+j_{3}^{2})}{2}
=j_{2}j_{3}. 
\end{equation*}

Therefore, the equation $w^{2}-(j_{1}+j_{4})w+j_{1}j_{4}=0$ has $4$ distinct roots 
$j_{1}$, $j_{2}$, $j_{3}$, and $j_{4}$, a contradiction.
\end{IEEEproof}

To sum up, in this subcase, $\mathbf{F}(\mathbf{X})-\mathbf{F}(\mathbf{Y})=\mathbf{F}=0^{n+2}$ which implies 
$\mathbf{x}=\mathbf{y}$. 

\begin{remark}
For $L^{\ast}(\mathbf{x},\mathbf{y})\leq 4$, 
the advantage of using Lemma \ref{M.l2.1} lies in the capability to separate errors into $4$ 
relatively independent parts, thus facilitating clearer discussions. 
This technique simultaneously entails the drawback of losing the first three conditions 
in \eqref{M.e3.1} for $\mathbf{U}$ and $\mathbf{V}$, whereas the first three conditions 
in \eqref{M.e3.1} are essential to derive 
$\sigma(\mathbf{F}(\mathbf{X})-\mathbf{F}(\mathbf{Y}))\leq 3$ in this subcase. 
For instance, when $\mathbf{U}=0111001110$ and $\mathbf{V}=0010000100$ fail to meet 
the first three conditions in \eqref{M.e3.1}, it results in 
$\sigma(\mathbf{F}(\mathbf{U})-\mathbf{F}(\mathbf{V}))=4$. 
Therefore, in this subcase we directly examine the check equations 
without employing Lemma \ref{M.l2.1} to complete the proof. 
\end{remark}




\subsection{One Substitution of Type Value $2$, One Substitution of Type Value $-2$, 
and Two Substitutions of Type Value $0$}\label{sec4.2}

Without loss of generality, we may assume the substitution of type value $2$ and the 
substitution of type value $-2$ occur at $X_{j_{k}}$ and $X_{j_{l}}$ respectively where $1\leq k<l\leq 4$. 
The crucial aspect of addressing this subcase lies in observing the following fact. 

\begin{lemma}\label{M.l4.2}
$\sigma(f_{j_{k}},f_{j_{k}+1},\cdots,f_{j_{l}})=1.$
\end{lemma}

\begin{IEEEproof}
At this moment, $f_{j_{k}}=1$, $f_{j_{l}}=1$, and $f_{i}\geq 1$ for $j_{k}<i<j_{l}$, 
which implies this lemma. 
\end{IEEEproof}

Regarding $f_{i}$ where $i\notin [j_{k},j_{l}]$, 
at most two of them are non-zero, corresponding to the positions of two substitutions of type value $0$. 
With the help of Lemma \ref{M.l4.2}, $\sigma(\mathbf{F})\leq 3$, which implies $\mathbf{x}=\mathbf{y}$. 




\subsection{Two Substitutions of Type Value $2$ and Two Substitutions of Type Value $-2$}\label{sec4.3}

Without loss of generality, we may assume the substitution occurring at $X_{j_{1}}$ is 
of type value $2$. Hence, it suffices to verify all possible type values of 
$(\mathbf{X},\mathbf{Y})$ below. 

\begin{itemize}

\item{$(2,2,-2,-2)$. At this moment, given that $\mathbf{F}$ is non-negative, $\sigma(\mathbf{F})=1\leq 3$.}

\item{$(2,-2,2,-2)$. At this moment, given that $\mathbf{F}$ is non-negative, $\sigma(\mathbf{F})=1\leq 3$.}

\item{$(2,-2,-2,2)$. At this moment, $f_{i}\geq 0$ for $i\in [1,j_{2}]$ and $f_{i}\leq 0$ for 
$i\in [j_{2}+1,n+2]$. Thus 
$\sigma(\mathbf{F})\leq \sigma((f_{1},\cdots,f_{j_{2}}))+\sigma((f_{j_{2}+1},\cdots,f_{n+2}))=2\leq 3$.}

\end{itemize}

To sum up, in this subcase, $\sigma(\mathbf{F})\leq 3$ which implies $\mathbf{x}=\mathbf{y}$. 




\section{$\mathcal{C}_{k_{1},k_{2},k_{3},k_{4}}$ is a $2$-Del Correcting Code}\label{sec5}

In this section, we assume the conditions in Theorem \ref{M.t3.1} hold, 
and proceed to verify $\mathbf{x}=\mathbf{y}$ 
in the case of $\mathcal{B}_{0,2,0}(\mathbf{x})\cap \mathcal{B}_{0,2,0}(\mathbf{y})\neq \varnothing$. 

Using Lemma \ref{M.l2.1}, we always assume 
$\mathbf{u},\mathbf{v}\in A_{2}^{m}$ such that $f(\mathbf{U})=f(\mathbf{V})$ and 
$\sigma(\mathbf{F}(\mathbf{X})-\mathbf{F}(\mathbf{Y}))\leq 
\sigma(\mathbf{F}(\mathbf{U})-\mathbf{F}(\mathbf{V}))$. 
Moreover, there exist $i_{1},i_{2},i_{3},i_{4}\in [2,m+1]$ with pairwise distances at least $5$, 
such that the sequence obtained by deleting $U_{i_{1}}$ and $U_{i_{2}}$ from $\mathbf{U}$ 
is equal to the sequence obtained by deleting $V_{i_{3}}$ and $V_{i_{4}}$ from $\mathbf{V}$. 
We arrange $i_{1},i_{2},i_{3},i_{4}$ in ascending order as $j_{1}<j_{2}<j_{3}<j_{4}$. 
For the sake of simplicity, in the rest of this section, 
$\mathbf{F}$ and $f_{i}$ denote $\mathbf{F}(\mathbf{U})-\mathbf{F}(\mathbf{V})$ 
and $f(\mathbf{U}_{[i]})-f(\mathbf{V}_{[i]})$, respectively. 

We may assume a deletion of type $\overline{del}$ occurs at $U_{j_{1}}$. Otherwise, 
given that $\sigma(\mathbf{F}(\mathbf{V})-\mathbf{F}(\mathbf{U}))
=\sigma(\mathbf{F}(\mathbf{U})-\mathbf{F}(\mathbf{V}))$, it suffices to 
consider the type of $(\mathbf{V},\mathbf{U})$ rather than $(\mathbf{U},\mathbf{V})$. 
According to this assumption, we divide our discussion into three subcases by the type of 
$(\mathbf{U},\mathbf{V})$. 




\subsection{The Type of $(\mathbf{U},\mathbf{V})$ is 
$(\overline{del},\overline{del},\underline{del},\underline{del})$}\label{sec5.1}

\begin{figure*}[htbp]
\centering
\begin{tikzpicture}

\node at (-0.5,1.5) {\small Position:};

\node at (0.5,1.5) {\tiny $1$};

\node at (1,1.5) {\tiny $2$};

\node at (2,1.5) {\tiny $j_{1}$$-$$2$};

\node at (3,1.5) {\tiny $j_{1}$};

\node at (4,1.5) {\tiny $j_{1}$$+$$2$};

\node at (5.5,1.5) {\tiny $j_{2}$$-$$1$};

\node at (6.5,1.5) {\tiny $j_{2}$$+$$1$};

\node at (8,1.5) {\tiny $j_{3}$$-$$2$};

\node at (9,1.5) {\tiny $j_{3}$};

\node at (10,1.5) {\tiny $j_{3}$$+$$2$};

\node at (11,1.5) {\tiny $j_{4}$$-$$2$};

\node at (12,1.5) {\tiny $j_{4}$};

\node at (13,1.5) {\tiny $j_{4}$$+$$2$};

\node at (14,1.5) {\tiny $m$$+$$2$};

\node at (-0.5,-1.5) {\small Position:};

\node at (0.5,-1.5) {\tiny $1$};

\node at (1,-1.5) {\tiny $2$};

\node at (2.5,-1.5) {\tiny $j_{1}$$-$$1$};

\node at (3.5,-1.5) {\tiny $j_{1}$$+$$1$};

\node at (5,-1.5) {\tiny $j_{2}$$-$$2$};

\node at (6,-1.5) {\tiny $j_{2}$};

\node at (7,-1.5) {\tiny $j_{2}$$+$$2$};

\node at (8.5,-1.5) {\tiny $j_{3}$$-$$1$};

\node at (9.5,-1.5) {\tiny $j_{3}$$+$$1$};

\node at (11.5,-1.5) {\tiny $j_{4}$$-$$1$};

\node at (12.5,-1.5) {\tiny $j_{4}$$+$$1$};

\node at (14,-1.5) {\tiny $m$$+$$2$};

\node[rotate=180] at (4.3,2) {$\underbrace{\hspace{2.8cm}}$};
\node[above] at (4.3,2) {\small $\{ e_{1},e_{1}-1 \}$};

\node at (7.85,-2) {$\underbrace{\hspace{3.8cm}}$};
\node[below] at (7.85,-2) {\small $\{ e_{1}+e_{2},e_{1}+e_{2}-1,e_{1}+e_{2}-2 \}$};

\node[rotate=180] at (10.95,2) {$\underbrace{\hspace{2.3cm}}$};
\node[above] at (10.95,2) {\small $\{ e_{1}+e_{2}+e_{3},e_{1}+e_{2}+e_{3}-1 \}$};

\node at (0,0.5) {\small$\mathbf{U}=$};

\fill (0.5,0.5) circle (2pt);

\fill (1,0.5) circle (2pt);

\node at (1.5,0.5) {\small $\cdots$};

\fill (2,0.5) circle (2pt);

\fill (2.5,0.5) circle (2pt);

\fill[red] (3,0.5) circle (2pt);

\fill (3.5,0.5) circle (2pt);

\fill (4,0.5) circle (2pt);

\node at (4.5,0.5) {\small $\cdots$};

\fill (5,0.5) circle (2pt);

\fill (5.5,0.5) circle (2pt);

\fill[red] (6,0.5) circle (2pt);

\fill (6.5,0.5) circle (2pt);

\fill (7,0.5) circle (2pt);

\node at (7.5,0.5) {\small $\cdots$};

\fill (8,0.5) circle (2pt);

\fill (8.5,0.5) circle (2pt);

\fill (9,0.5) circle (2pt);

\fill (9.5,0.5) circle (2pt);

\fill (10,0.5) circle (2pt);

\node at (10.5,0.5) {\small $\cdots$};

\fill (11,0.5) circle (2pt);

\fill (11.5,0.5) circle (2pt);

\fill (12,0.5) circle (2pt);

\fill (12.5,0.5) circle (2pt);

\fill (13,0.5) circle (2pt);

\node at (13.5,0.5) {\small $\cdots$};

\fill (14,0.5) circle (2pt);

\node at (0,-0.5) {\small$\mathbf{V}=$};

\fill (0.5,-0.5) circle (2pt);

\fill (1,-0.5) circle (2pt);

\node at (1.5,-0.5) {\small $\cdots$};

\fill (2,-0.5) circle (2pt);

\fill (2.5,-0.5) circle (2pt);

\fill (3,-0.5) circle (2pt);

\fill (3.5,-0.5) circle (2pt);

\fill (4,-0.5) circle (2pt);

\node at (4.5,-0.5) {\small $\cdots$};

\fill (5,-0.5) circle (2pt);

\fill (5.5,-0.5) circle (2pt);

\fill (6,-0.5) circle (2pt);

\fill (6.5,-0.5) circle (2pt);

\fill (7,-0.5) circle (2pt);

\node at (7.5,-0.5) {\small $\cdots$};

\fill (8,-0.5) circle (2pt);

\fill (8.5,-0.5) circle (2pt);

\fill[red] (9,-0.5) circle (2pt);

\fill (9.5,-0.5) circle (2pt);

\fill (10,-0.5) circle (2pt);

\node at (10.5,-0.5) {\small $\cdots$};

\fill (11,-0.5) circle (2pt);

\fill (11.5,-0.5) circle (2pt);

\fill[red] (12,-0.5) circle (2pt);

\fill (12.5,-0.5) circle (2pt);

\fill (13,-0.5) circle (2pt);

\node at (13.5,-0.5) {\small $\cdots$};

\fill (14,-0.5) circle (2pt);

\draw (0.5,0.5)--(0.5,-0.5);

\draw (1,0.5)--(1,-0.5);

\draw (2,0.5)--(2,-0.5);

\draw (2.5,0.5)--(2.5,-0.5);

\draw (3.5,0.5)--(3,-0.5);

\draw (4,0.5)--(3.5,-0.5);

\draw (5.5,0.5)--(5,-0.5);

\draw (6.5,0.5)--(5.5,-0.5);

\draw (7,0.5)--(6,-0.5);

\draw (9,0.5)--(8,-0.5);

\draw (9.5,0.5)--(8.5,-0.5);

\draw (10,0.5)--(9.5,-0.5);

\draw (11.5,0.5)--(11,-0.5);

\draw (12,0.5)--(11.5,-0.5);

\draw (12.5,0.5)--(12.5,-0.5);

\draw (13,0.5)--(13,-0.5);

\draw (14,0.5)--(14,-0.5);

\end{tikzpicture}
\caption{The type of $(\mathbf{U},\mathbf{V})$ is 
$(\overline{del},\overline{del},\underline{del},\underline{del})$.}
\label{M.f5.a}
\end{figure*}
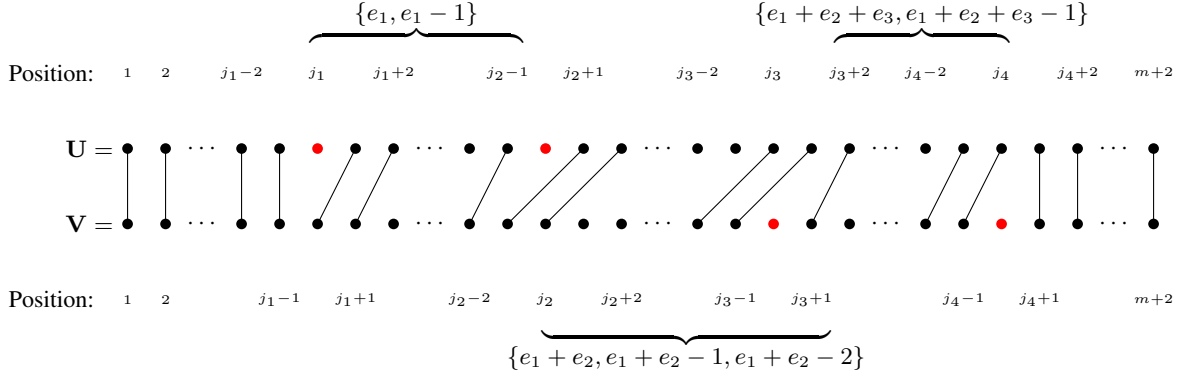

Assume the type value of $(\mathbf{U},\mathbf{V})$ is 
$(\overline{e_{1}},\overline{e_{2}},\underline{e_{3}},\underline{e_{4}})$. 
That is to say, deletions of type value $\overline{e_{1}}$, $\overline{e_{2}}$, 
$\underline{e_{3}}$, and $\underline{e_{4}}$ 
occur at $U_{j_{1}}$, $U_{j_{2}}$, $V_{j_{3}}$, and $V_{j_{4}}$, respectively. 
Referring to Fig. \ref{M.f5.a}, 
$e_{1}=f(\mathbf{U}_{[j_{1}-1,j_{1}+1]})-f(\mathbf{V}_{[j_{1}-1,j_{1}]})$, 
$e_{2}=f(\mathbf{U}_{[j_{2}-1,j_{2}+1]})-f(\mathbf{V}_{[j_{2}-2,j_{2}-1]})$, 
$e_{3}=f(\mathbf{U}_{[j_{3}+1,j_{3}+2]})-f(\mathbf{V}_{[j_{3}-1,j_{3}+1]})$, 
$e_{4}=f(\mathbf{U}_{[j_{4},j_{4}+1]})-f(\mathbf{V}_{[j_{4}-1,j_{4}+1]})$. 
We discuss $f_{i}$ for $i\in [m+2]$ in this subcase. 

\begin{itemize}
\item For $i\in [1,j_{1}-1]$, $f_{i}=0$.

\item For $i=j_{1}$,
\begin{equation*}
\begin{split}
f_{i}=&f(\mathbf{U}_{[j_{1}-1,j_{1}]})-f(\mathbf{V}_{[j_{1}-1,j_{1}]})
\\
=&f(\mathbf{U}_{[j_{1}-1,j_{1}+1]})-f(\mathbf{V}_{[j_{1}-1,j_{1}]})-f(\mathbf{U}_{[j_{1},j_{1}+1]})
\\
=&e_{1}-f(\mathbf{U}_{[j_{1},j_{1}+1]})
\\
\in&\{ e_{1},e_{1}-1 \}.
\end{split}
\end{equation*}

\item For $i\in [j_{1}+1,j_{2}-1]$, 
\begin{equation*}
\begin{split}
f_{i}=e_{1}-f(\mathbf{V}_{[i-1,i]})\in\{ e_{1},e_{1}-1 \}.
\end{split}
\end{equation*}

\item For $i=j_{2}$, 
\begin{equation*}
\begin{split}
f_{i}=&e_{1}+f(\mathbf{U}_{[j_{2}-1,j_{2}]})-f(\mathbf{V}_{[j_{1}-2,j_{2}-1]})
\\
&-f(\mathbf{V}_{[j_{2}-1,j_{2}]})
\\
=&e_{1}+f(\mathbf{U}_{[j_{2}-1,j_{2}+1]})-f(\mathbf{U}_{[j_{2},j_{2}+1]})
\\
&-f(\mathbf{V}_{[j_{2}-2,j_{2}-1]})-f(\mathbf{V}_{[j_{2}-1,j_{2}]})
\\
=&e_{1}+e_{2}-f(\mathbf{U}_{[j_{2},j_{2}+1]})
-f(\mathbf{V}_{[j_{2}-1,j_{2}]})
\\
\in&\{ e_{1}+e_{2},e_{1}+e_{2}-1,e_{1}+e_{2}-2 \}.
\end{split}
\end{equation*}

\item For $i\in [j_{2}+1,j_{3}+1]$, 
\begin{equation*}
\begin{split}
f_{i}=&e_{1}+e_{2}-f(\mathbf{V}_{[i-2,i-1]})-f(\mathbf{V}_{[i-1,i]})
\\
\in&\{ e_{1}+e_{2},e_{1}+e_{2}-1,e_{1}+e_{2}-2 \}.
\end{split}
\end{equation*}

\item For $i\in [j_{3}+2,j_{4}]$, 
\begin{equation*}
\begin{split}
f_{i}=&e_{1}+e_{2}+e_{3}-f(\mathbf{V}_{[i-1,i]})
\\
\in&\{ e_{1}+e_{2}+e_{3},e_{1}+e_{2}+e_{3}-1 \}.
\end{split}
\end{equation*}

\item For $i\in [j_{4}+1,m+2]$, $f_{i}=0$.

\end{itemize}

Noting that $e_{1},e_{2},e_{3}\in\{ 2,0,-2 \}$, we obtain 
\begin{equation*}
\begin{split}
&\sigma(\mathbf{F}(\mathbf{X})-\mathbf{F}(\mathbf{Y}))
\\
\leq&\sigma(\mathbf{F}(\mathbf{U})-\mathbf{F}(\mathbf{V}))
\\
\leq&\sigma((f_{1},\cdots,f_{j_{2}-1}))+\sigma((f_{j_{2}},\cdots,f_{j_{3}+1}))
\\
&+\sigma((f_{j_{3}+2},\cdots,f_{m+2}))
\\
=&3,
\end{split}
\end{equation*}
which implies $\mathbf{x}=\mathbf{y}$. 




\subsection{The Type of $(\mathbf{U},\mathbf{V})$ is 
$(\overline{del},\underline{del},\overline{del},\underline{del})$}\label{sec5.2}

\begin{figure*}[htbp]
\centering
\begin{tikzpicture}

\node at (-0.5,1.5) {\small Position:};

\node at (0.5,1.5) {\tiny $1$};

\node at (1,1.5) {\tiny $2$};

\node at (2,1.5) {\tiny $j_{1}$$-$$2$};

\node at (3,1.5) {\tiny $j_{1}$};

\node at (4,1.5) {\tiny $j_{1}$$+$$2$};

\node at (5,1.5) {\tiny $j_{2}$$-$$2$};

\node at (6,1.5) {\tiny $j_{2}$};

\node at (7,1.5) {\tiny $j_{2}$$+$$2$};

\node at (8,1.5) {\tiny $j_{3}$$-$$2$};

\node at (9,1.5) {\tiny $j_{3}$};

\node at (10,1.5) {\tiny $j_{3}$$+$$2$};

\node at (11,1.5) {\tiny $j_{4}$$-$$2$};

\node at (12,1.5) {\tiny $j_{4}$};

\node at (13,1.5) {\tiny $j_{4}$$+$$2$};

\node at (14,1.5) {\tiny $m$$+$$2$};

\node at (-0.5,-1.5) {\small Position:};

\node at (0.5,-1.5) {\tiny $1$};

\node at (1,-1.5) {\tiny $2$};

\node at (2.5,-1.5) {\tiny $j_{1}$$-$$1$};

\node at (3.5,-1.5) {\tiny $j_{1}$$+$$1$};

\node at (5.5,-1.5) {\tiny $j_{2}$$-$$1$};

\node at (6.5,-1.5) {\tiny $j_{2}$$+$$1$};

\node at (8.5,-1.5) {\tiny $j_{3}$$-$$1$};

\node at (9.5,-1.5) {\tiny $j_{3}$$+$$1$};

\node at (11.5,-1.5) {\tiny $j_{4}$$-$$1$};

\node at (12.5,-1.5) {\tiny $j_{4}$$+$$1$};

\node at (14,-1.5) {\tiny $m$$+$$2$};

\node[rotate=180] at (4.5,2) {$\underbrace{\hspace{3.1cm}}$};
\node[above] at (4.5,2) {\small $\{ e_{1},e_{1}-1 \}$};

\node at (7.5,-2) {$\underbrace{\hspace{2.5cm}}$};
\node[below] at (7.5,-2) {\small $\{ e_{1}+e_{2}\}$};

\node[rotate=180] at (10.5,2) {$\underbrace{\hspace{3.1cm}}$};
\node[above] at (10.5,2) {\small $\{ e_{1}+e_{2}+e_{3},e_{1}+e_{2}+e_{3}-1 \}$};

\node at (0,0.5) {\small$\mathbf{U}=$};

\fill (0.5,0.5) circle (2pt);

\fill (1,0.5) circle (2pt);

\node at (1.5,0.5) {\small $\cdots$};

\fill (2,0.5) circle (2pt);

\fill (2.5,0.5) circle (2pt);

\fill[red] (3,0.5) circle (2pt);

\fill (3.5,0.5) circle (2pt);

\fill (4,0.5) circle (2pt);

\node at (4.5,0.5) {\small $\cdots$};

\fill (5,0.5) circle (2pt);

\fill (5.5,0.5) circle (2pt);

\fill (6,0.5) circle (2pt);

\fill (6.5,0.5) circle (2pt);

\fill (7,0.5) circle (2pt);

\node at (7.5,0.5) {\small $\cdots$};

\fill (8,0.5) circle (2pt);

\fill (8.5,0.5) circle (2pt);

\fill[red] (9,0.5) circle (2pt);

\fill (9.5,0.5) circle (2pt);

\fill (10,0.5) circle (2pt);

\node at (10.5,0.5) {\small $\cdots$};

\fill (11,0.5) circle (2pt);

\fill (11.5,0.5) circle (2pt);

\fill (12,0.5) circle (2pt);

\fill (12.5,0.5) circle (2pt);

\fill (13,0.5) circle (2pt);

\node at (13.5,0.5) {\small $\cdots$};

\fill (14,0.5) circle (2pt);

\node at (0,-0.5) {\small$\mathbf{V}=$};

\fill (0.5,-0.5) circle (2pt);

\fill (1,-0.5) circle (2pt);

\node at (1.5,-0.5) {\small $\cdots$};

\fill (2,-0.5) circle (2pt);

\fill (2.5,-0.5) circle (2pt);

\fill (3,-0.5) circle (2pt);

\fill (3.5,-0.5) circle (2pt);

\fill (4,-0.5) circle (2pt);

\node at (4.5,-0.5) {\small $\cdots$};

\fill (5,-0.5) circle (2pt);

\fill (5.5,-0.5) circle (2pt);

\fill[red] (6,-0.5) circle (2pt);

\fill (6.5,-0.5) circle (2pt);

\fill (7,-0.5) circle (2pt);

\node at (7.5,-0.5) {\small $\cdots$};

\fill (8,-0.5) circle (2pt);

\fill (8.5,-0.5) circle (2pt);

\fill (9,-0.5) circle (2pt);

\fill (9.5,-0.5) circle (2pt);

\fill (10,-0.5) circle (2pt);

\node at (10.5,-0.5) {\small $\cdots$};

\fill (11,-0.5) circle (2pt);

\fill (11.5,-0.5) circle (2pt);

\fill[red] (12,-0.5) circle (2pt);

\fill (12.5,-0.5) circle (2pt);

\fill (13,-0.5) circle (2pt);

\node at (13.5,-0.5) {\small $\cdots$};

\fill (14,-0.5) circle (2pt);

\draw (0.5,0.5)--(0.5,-0.5);

\draw (1,0.5)--(1,-0.5);

\draw (2,0.5)--(2,-0.5);

\draw (2.5,0.5)--(2.5,-0.5);

\draw (3.5,0.5)--(3,-0.5);

\draw (4,0.5)--(3.5,-0.5);

\draw (5.5,0.5)--(5,-0.5);

\draw (6,0.5)--(5.5,-0.5);

\draw (6.5,0.5)--(6.5,-0.5);

\draw (7,0.5)--(7,-0.5);

\draw (8,0.5)--(8,-0.5);

\draw (8.5,0.5)--(8.5,-0.5);

\draw (9.5,0.5)--(9,-0.5);

\draw (10,0.5)--(9.5,-0.5);

\draw (11.5,0.5)--(11,-0.5);

\draw (12,0.5)--(11.5,-0.5);

\draw (12.5,0.5)--(12.5,-0.5);

\draw (13,0.5)--(13,-0.5);

\draw (14,0.5)--(14,-0.5);

\end{tikzpicture}
\caption{The type of $(\mathbf{U},\mathbf{V})$ is 
$(\overline{del},\underline{del},\overline{del},\underline{del})$.}
\label{M.f5.b}
\end{figure*}
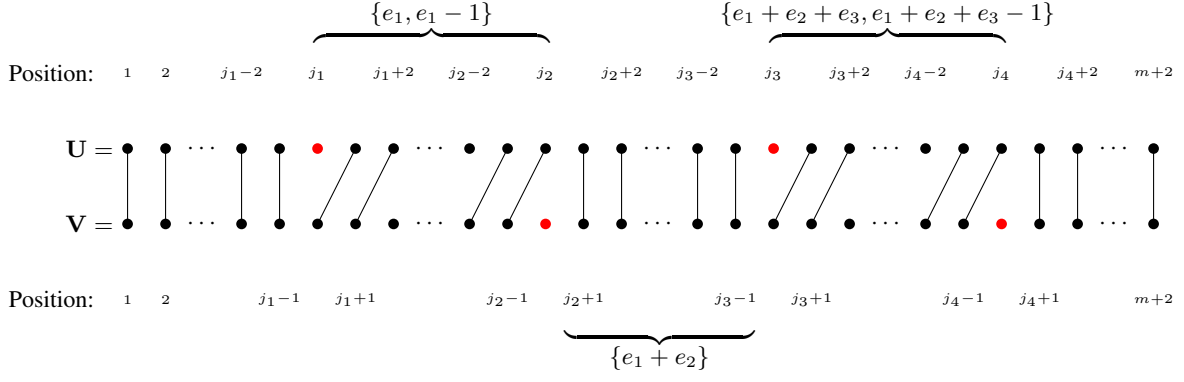

Assume the type value of $(\mathbf{U},\mathbf{V})$ is 
$(\overline{e_{1}},\underline{e_{2}},\overline{e_{3}},\underline{e_{4}})$. 
That is to say, deletions of type value $\overline{e_{1}}$, $\underline{e_{2}}$, 
$\overline{e_{3}}$, and $\underline{e_{4}}$ 
occur at $U_{j_{1}}$, $V_{j_{2}}$, $U_{j_{3}}$, and $V_{j_{4}}$, respectively. 
Referring to Fig. \ref{M.f5.b}, $e_{1}=f(\mathbf{U}_{[j_{1}-1,j_{1}+1]})-f(\mathbf{V}_{[j_{1}-1,j_{1}]})$, 
$e_{2}=f(\mathbf{U}_{[j_{2},j_{2}+1]})-f(\mathbf{V}_{[j_{2}-1,j_{2}+1]})$, 
$e_{3}=f(\mathbf{U}_{[j_{3}-1,j_{3}+1]})-f(\mathbf{V}_{[j_{3}-1,j_{3}]})$, 
$e_{4}=f(\mathbf{U}_{[j_{4},j_{4}+1]})-f(\mathbf{V}_{[j_{4}-1,j_{4}+1]})$. 
We discuss $f_{i}$ for $i\in [m+2]$ in this subcase. 

\begin{itemize}
\item For $i\in [1,j_{1}-1]$, $f_{i}=0$.

\item For $i=j_{1}$,
\begin{equation*}
\begin{split}
f_{i}=&f(\mathbf{U}_{[j_{1}-1,j_{1}]})-f(\mathbf{V}_{[j_{1}-1,j_{1}]})
\\
=&f(\mathbf{U}_{[j_{1}-1,j_{1}+1]})-f(\mathbf{V}_{[j_{1}-1,j_{1}]})-f(\mathbf{U}_{[j_{1},j_{1}+1]})
\\
=&e_{1}-f(\mathbf{U}_{[j_{1},j_{1}+1]})
\\
\in&\{ e_{1},e_{1}-1 \}.
\end{split}
\end{equation*}

\item For $i\in [j_{1}+1,j_{2}]$, 
\begin{equation*}
\begin{split}
f_{i}=e_{1}-f(\mathbf{V}_{[i-1,i]})\in\{ e_{1},e_{1}-1 \}.
\end{split}
\end{equation*}

\item For $i\in [j_{2}+1,j_{3}-1]$, $f_{i}=e_{1}+e_{2}$.

\item For $i=j_{3}$, 
\begin{equation*}
\begin{split}
f_{i}=&e_{1}+e_{2}+f(\mathbf{U}_{[j_{3}-1,j_{3}]})-f(\mathbf{V}_{[j_{3}-1,j_{3}]})
\\
=&e_{1}+e_{2}+f(\mathbf{U}_{[j_{3}-1,j_{3}+1]})-f(\mathbf{V}_{[j_{3}-1,j_{3}]})
\\
&-f(\mathbf{U}_{[j_{3},j_{3}+1]})
\\
=&e_{1}+e_{2}+e_{3}-f(\mathbf{U}_{[j_{3},j_{3}+1]})
\\
\in&\{ e_{1}+e_{2}+e_{3},e_{1}+e_{2}+e_{3}-1 \}.
\end{split}
\end{equation*}

\item For $i\in [j_{3}+1,j_{4}]$, 
\begin{equation*}
\begin{split}
f_{i}=&e_{1}+e_{2}+e_{3}-f(\mathbf{V}_{[i-1,i]})
\\
\in&\{ e_{1}+e_{2}+e_{3},e_{1}+e_{2}+e_{3}-1 \}.
\end{split}
\end{equation*}

\item For $i\in [j_{4}+1,m+2]$, $f_{i}=0$.

\end{itemize}

Noting that $e_{1},e_{2},e_{3}\in\{ 2,0,-2 \}$, we obtain 
\begin{equation*}
\begin{split}
&\sigma(\mathbf{F}(\mathbf{X})-\mathbf{F}(\mathbf{Y}))
\\
\leq&\sigma(\mathbf{F}(\mathbf{U})-\mathbf{F}(\mathbf{V}))
\\
\leq&\sigma((f_{1},\cdots,f_{j_{2}}))+\sigma((f_{j_{2}+1},\cdots,f_{j_{3}-1}))
\\
&+\sigma((f_{j_{3}},\cdots,f_{m+2}))
\\
=&3,
\end{split}
\end{equation*}
which implies $\mathbf{x}=\mathbf{y}$. 




\subsection{The Type of $(\mathbf{U},\mathbf{V})$ is 
$(\overline{del},\underline{del},\underline{del},\overline{del})$}\label{sec5.3}

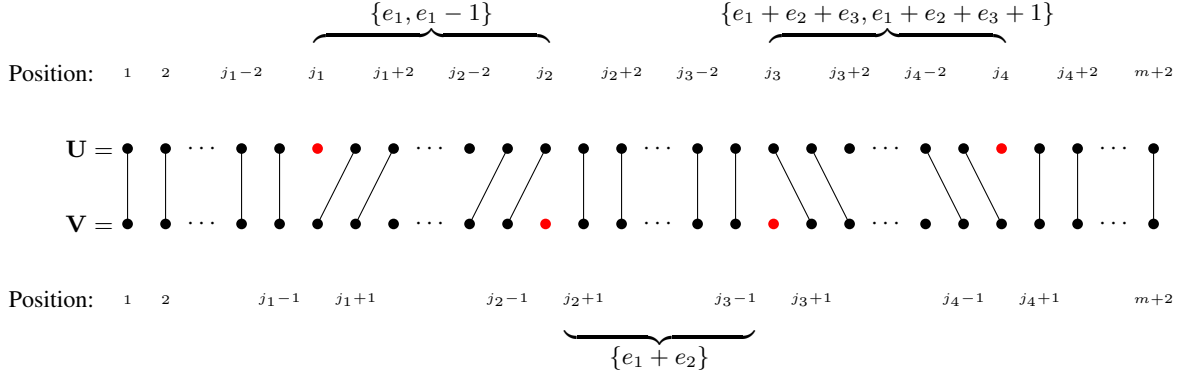
\begin{figure*}[htbp]
\centering
\begin{tikzpicture}

\node at (-0.5,1.5) {\small Position:};

\node at (0.5,1.5) {\tiny $1$};

\node at (1,1.5) {\tiny $2$};

\node at (2,1.5) {\tiny $j_{1}$$-$$2$};

\node at (3,1.5) {\tiny $j_{1}$};

\node at (4,1.5) {\tiny $j_{1}$$+$$2$};

\node at (5,1.5) {\tiny $j_{2}$$-$$2$};

\node at (6,1.5) {\tiny $j_{2}$};

\node at (7,1.5) {\tiny $j_{2}$$+$$2$};

\node at (8,1.5) {\tiny $j_{3}$$-$$2$};

\node at (9,1.5) {\tiny $j_{3}$};

\node at (10,1.5) {\tiny $j_{3}$$+$$2$};

\node at (11,1.5) {\tiny $j_{4}$$-$$2$};

\node at (12,1.5) {\tiny $j_{4}$};

\node at (13,1.5) {\tiny $j_{4}$$+$$2$};

\node at (14,1.5) {\tiny $m$$+$$2$};

\node at (-0.5,-1.5) {\small Position:};

\node at (0.5,-1.5) {\tiny $1$};

\node at (1,-1.5) {\tiny $2$};

\node at (2.5,-1.5) {\tiny $j_{1}$$-$$1$};

\node at (3.5,-1.5) {\tiny $j_{1}$$+$$1$};

\node at (5.5,-1.5) {\tiny $j_{2}$$-$$1$};

\node at (6.5,-1.5) {\tiny $j_{2}$$+$$1$};

\node at (8.5,-1.5) {\tiny $j_{3}$$-$$1$};

\node at (9.5,-1.5) {\tiny $j_{3}$$+$$1$};

\node at (11.5,-1.5) {\tiny $j_{4}$$-$$1$};

\node at (12.5,-1.5) {\tiny $j_{4}$$+$$1$};

\node at (14,-1.5) {\tiny $m$$+$$2$};

\node[rotate=180] at (4.5,2) {$\underbrace{\hspace{3.1cm}}$};
\node[above] at (4.5,2) {\small $\{ e_{1},e_{1}-1 \}$};

\node at (7.5,-2) {$\underbrace{\hspace{2.5cm}}$};
\node[below] at (7.5,-2) {\small $\{ e_{1}+e_{2}\}$};

\node[rotate=180] at (10.5,2) {$\underbrace{\hspace{3.1cm}}$};
\node[above] at (10.5,2) {\small $\{ e_{1}+e_{2}+e_{3},e_{1}+e_{2}+e_{3}+1 \}$};

\node at (0,0.5) {\small$\mathbf{U}=$};

\fill (0.5,0.5) circle (2pt);

\fill (1,0.5) circle (2pt);

\node at (1.5,0.5) {\small $\cdots$};

\fill (2,0.5) circle (2pt);

\fill (2.5,0.5) circle (2pt);

\fill[red] (3,0.5) circle (2pt);

\fill (3.5,0.5) circle (2pt);

\fill (4,0.5) circle (2pt);

\node at (4.5,0.5) {\small $\cdots$};

\fill (5,0.5) circle (2pt);

\fill (5.5,0.5) circle (2pt);

\fill (6,0.5) circle (2pt);

\fill (6.5,0.5) circle (2pt);

\fill (7,0.5) circle (2pt);

\node at (7.5,0.5) {\small $\cdots$};

\fill (8,0.5) circle (2pt);

\fill (8.5,0.5) circle (2pt);

\fill (9,0.5) circle (2pt);

\fill (9.5,0.5) circle (2pt);

\fill (10,0.5) circle (2pt);

\node at (10.5,0.5) {\small $\cdots$};

\fill (11,0.5) circle (2pt);

\fill (11.5,0.5) circle (2pt);

\fill[red] (12,0.5) circle (2pt);

\fill (12.5,0.5) circle (2pt);

\fill (13,0.5) circle (2pt);

\node at (13.5,0.5) {\small $\cdots$};

\fill (14,0.5) circle (2pt);

\node at (0,-0.5) {\small$\mathbf{V}=$};

\fill (0.5,-0.5) circle (2pt);

\fill (1,-0.5) circle (2pt);

\node at (1.5,-0.5) {\small $\cdots$};

\fill (2,-0.5) circle (2pt);

\fill (2.5,-0.5) circle (2pt);

\fill (3,-0.5) circle (2pt);

\fill (3.5,-0.5) circle (2pt);

\fill (4,-0.5) circle (2pt);

\node at (4.5,-0.5) {\small $\cdots$};

\fill (5,-0.5) circle (2pt);

\fill (5.5,-0.5) circle (2pt);

\fill[red] (6,-0.5) circle (2pt);

\fill (6.5,-0.5) circle (2pt);

\fill (7,-0.5) circle (2pt);

\node at (7.5,-0.5) {\small $\cdots$};

\fill (8,-0.5) circle (2pt);

\fill (8.5,-0.5) circle (2pt);

\fill[red] (9,-0.5) circle (2pt);

\fill (9.5,-0.5) circle (2pt);

\fill (10,-0.5) circle (2pt);

\node at (10.5,-0.5) {\small $\cdots$};

\fill (11,-0.5) circle (2pt);

\fill (11.5,-0.5) circle (2pt);

\fill (12,-0.5) circle (2pt);

\fill (12.5,-0.5) circle (2pt);

\fill (13,-0.5) circle (2pt);

\node at (13.5,-0.5) {\small $\cdots$};

\fill (14,-0.5) circle (2pt);

\draw (0.5,0.5)--(0.5,-0.5);

\draw (1,0.5)--(1,-0.5);

\draw (2,0.5)--(2,-0.5);

\draw (2.5,0.5)--(2.5,-0.5);

\draw (3.5,0.5)--(3,-0.5);

\draw (4,0.5)--(3.5,-0.5);

\draw (5.5,0.5)--(5,-0.5);

\draw (6,0.5)--(5.5,-0.5);

\draw (6.5,0.5)--(6.5,-0.5);

\draw (7,0.5)--(7,-0.5);

\draw (8,0.5)--(8,-0.5);

\draw (8.5,0.5)--(8.5,-0.5);

\draw (9,0.5)--(9.5,-0.5);

\draw (9.5,0.5)--(10,-0.5);

\draw (11,0.5)--(11.5,-0.5);

\draw (11.5,0.5)--(12,-0.5);

\draw (12.5,0.5)--(12.5,-0.5);

\draw (13,0.5)--(13,-0.5);

\draw (14,0.5)--(14,-0.5);

\end{tikzpicture}
\caption{The type of $(\mathbf{U},\mathbf{V})$ is 
$(\overline{del},\underline{del},\underline{del},\overline{del})$.}
\label{M.f5.c}
\end{figure*}

Assume the type value of $(\mathbf{U},\mathbf{V})$ is 
$(\overline{e_{1}},\underline{e_{2}},\underline{e_{3}},\overline{e_{4}})$. 
That is to say, deletions of type value $\overline{e_{1}}$, $\underline{e_{2}}$, 
$\underline{e_{3}}$, and $\overline{e_{4}}$ 
occur at $U_{j_{1}}$, $V_{j_{2}}$, $V_{j_{3}}$, and $U_{j_{4}}$, respectively. 
Referring to Fig. \ref{M.f5.c}, $e_{1}=f(\mathbf{U}_{[j_{1}-1,j_{1}+1]})-f(\mathbf{V}_{[j_{1}-1,j_{1}]})$, 
$e_{2}=f(\mathbf{U}_{[j_{2},j_{2}+1]})-f(\mathbf{V}_{[j_{2}-1,j_{2}+1]})$, 
$e_{3}=f(\mathbf{U}_{[j_{3}-1,j_{3}]})-f(\mathbf{V}_{[j_{3}-1,j_{3}+1]})$, 
$e_{4}=f(\mathbf{U}_{[j_{4}-1,j_{4}+1]})-f(\mathbf{V}_{[j_{4},j_{4}+1]})$. 
We discuss $f_{i}$ for $i\in [m+2]$ in this subcase. 

\begin{itemize}
\item For $i\in [1,j_{1}-1]$, $f_{i}=0$.

\item For $i=j_{1}$,
\begin{equation*}
\begin{split}
f_{i}=&f(\mathbf{U}_{[j_{1}-1,j_{1}]})-f(\mathbf{V}_{[j_{1}-1,j_{1}]})
\\
=&f(\mathbf{U}_{[j_{1}-1,j_{1}+1]})-f(\mathbf{V}_{[j_{1}-1,j_{1}]})-f(\mathbf{U}_{[j_{1},j_{1}+1]})
\\
=&e_{1}-f(\mathbf{U}_{[j_{1},j_{1}+1]})
\\
\in&\{ e_{1},e_{1}-1 \}.
\end{split}
\end{equation*}

\item For $i\in [j_{1}+1,j_{2}]$, 
\begin{equation*}
\begin{split}
f_{i}=e_{1}-f(\mathbf{V}_{[i-1,i]})\in\{ e_{1},e_{1}-1 \}.
\end{split}
\end{equation*}

\item For $i\in [j_{2}+1,j_{3}-1]$, $f_{i}=e_{1}+e_{2}$. 

\item For $i=j_{3}$, 
\begin{equation*}
\begin{split}
f_{i}=&e_{1}+e_{2}+f(\mathbf{U}_{[j_{3}-1,j_{3}]})-f(\mathbf{V}_{[j_{3}-1,j_{3}]})
\\
=&e_{1}+e_{2}+f(\mathbf{U}_{[j_{3}-1,j_{3}]})-f(\mathbf{V}_{[j_{3}-1,j_{3}+1]})
\\
&+f(\mathbf{V}_{[j_{3},j_{3}+1]})
\\
=&e_{1}+e_{2}+e_{3}+f(\mathbf{V}_{[j_{3},j_{3}+1]})
\\
\in&\{ e_{1}+e_{2}+e_{3},e_{1}+e_{2}+e_{3}+1 \}.
\end{split}
\end{equation*}

\item For $i\in [j_{3}+1,j_{4}]$, 
\begin{equation*}
\begin{split}
f_{i}=&e_{1}+e_{2}+e_{3}+f(\mathbf{U}_{[i-1,i]})
\\
\in&\{ e_{1}+e_{2}+e_{3},e_{1}+e_{2}+e_{3}+1 \}.
\end{split}
\end{equation*}

\item For $i\in [j_{4}+1,m+2]$, $f_{i}=0$.

\end{itemize}

Noting that $e_{1},e_{2},e_{3}\in\{ 2,0,-2 \}$, we obtain 
\begin{equation*}
\begin{split}
&\sigma(\mathbf{F}(\mathbf{X})-\mathbf{F}(\mathbf{Y}))
\\
\leq&\sigma(\mathbf{F}(\mathbf{U})-\mathbf{F}(\mathbf{V}))
\\
\leq&\sigma((f_{1},\cdots,f_{j_{2}}))+\sigma((f_{j_{2}+1},\cdots,f_{j_{3}-1}))
\\
&+\sigma((f_{j_{3}},\cdots,f_{m+2}))
\\
=&3,
\end{split}
\end{equation*}
which implies $\mathbf{x}=\mathbf{y}$. 




\section{$\mathcal{C}_{k_{1},k_{2},k_{3},k_{4}}$ is a $1$-Del $1$-Sub Correcting Code}\label{sec6}

In this section, we assume the conditions in Theorem \ref{M.t3.1} hold, 
and proceed to verify $\mathbf{x}=\mathbf{y}$ 
in the case of $\mathcal{B}_{0,1,1}(\mathbf{x})\cap \mathcal{B}_{0,1,1}(\mathbf{y})\neq \varnothing$. 

Using Lemma \ref{M.l2.1}, we always assume 
$\mathbf{u},\mathbf{v}\in A_{2}^{m}$ such that $f(\mathbf{U})=f(\mathbf{V})$ and 
$\sigma(\mathbf{F}(\mathbf{X})-\mathbf{F}(\mathbf{Y}))\leq 
\sigma(\mathbf{F}(\mathbf{U})-\mathbf{F}(\mathbf{V}))$. 
Moreover, there exist $i_{1},i_{2},i_{3},i_{4}\in [2,m+1]$ with pairwise distances at least $5$, 
such that the sequence obtained by deleting $U_{i_{1}}$ and substituting 
$U_{i_{2}}$ and $U_{i_{3}}$ (allowing trivial substitutions) from $\mathbf{U}$ 
is equal to the sequence obtained by deleting $V_{i_{4}}$ from $\mathbf{V}$. 
We arrange $i_{1},i_{2},i_{3},i_{4}$ in ascending order as $j_{1}<j_{2}<j_{3}<j_{4}$. 
For the sake of simplicity, in the rest of this section, $\mathbf{F}$ and $f_{i}$ denote 
$\mathbf{F}(\mathbf{U})-\mathbf{F}(\mathbf{V})$ and $f(\mathbf{U}_{[i]})-f(\mathbf{V}_{[i]})$, 
respectively. 

There are a total of $12$ possible types of $(\mathbf{U},\mathbf{V})$. 
A concept is introduced to simplify the discussion. 

\begin{definition}
The inversion of $\mathbf{z}\in \mathbb{Z}^{n}$ is defined as $\mathbf{z}^{-1}=z_{n}z_{n-1}\cdots z_{1}$. 
\end{definition}

By definition, $f(\mathbf{z})=f(\mathbf{z}^{-1})$, $(\mathbf{z}^{-1})_{[i]}=z_{n}z_{n-1}\cdots z_{n-i+1}$, 
while $(\mathbf{z}_{[i]})^{-1}=z_{i}z_{i-1}\cdots z_{1}$. 

\begin{lemma}\label{M.l6.1}
If $\mathbf{x},\mathbf{y}\in A_{2}^{n},f(\mathbf{x})=f(\mathbf{y})$, 
then $\sigma(\mathbf{F}(\mathbf{x}^{-1})-\mathbf{F}(\mathbf{y}^{-1}))
=\sigma(\mathbf{F}(\mathbf{x})-\mathbf{F}(\mathbf{y}))$.
\end{lemma}

\begin{IEEEproof}
\begin{equation*}
\begin{split}
&f((\mathbf{x}^{-1})_{[i]})-f((\mathbf{y}^{-1})_{[i]})
\\
=&f(x_{n}x_{n-1}\cdots x_{n-i+1})-f(y_{n}y_{n-1}\cdots y_{n-i+1})
\\
=&f(\mathbf{x}_{[n-i+1,n]})-f(\mathbf{y}_{[n-i+1,n]})
\\
=&(f(\mathbf{x})-f(\mathbf{x}_{[n-i+1]}))-(f(\mathbf{y})-f(\mathbf{y}_{[n-i+1]}))
\\
=&f(\mathbf{y}_{[n-i+1]})-f(\mathbf{x}_{[n-i+1]})
\\
=&-(f(\mathbf{x}_{[n-i+1]})-f(\mathbf{y}_{[n-i+1]})).
\end{split}
\end{equation*}

Therefore, $\mathbf{F}(\mathbf{x}^{-1})-\mathbf{F}(\mathbf{y}^{-1})
=-(\mathbf{F}(\mathbf{x})-\mathbf{F}(\mathbf{y}))^{-1}$ which implies 
\begin{equation*}
\begin{split}
\sigma(\mathbf{F}(\mathbf{x}^{-1})-\mathbf{F}(\mathbf{y}^{-1}))
=&\sigma(-(\mathbf{F}(\mathbf{x})-\mathbf{F}(\mathbf{y}))^{-1})
\\
=&\sigma((\mathbf{F}(\mathbf{x})-\mathbf{F}(\mathbf{y}))^{-1})
\\
=&\sigma(\mathbf{F}(\mathbf{x})-\mathbf{F}(\mathbf{y})).
\end{split}
\end{equation*}
\end{IEEEproof}

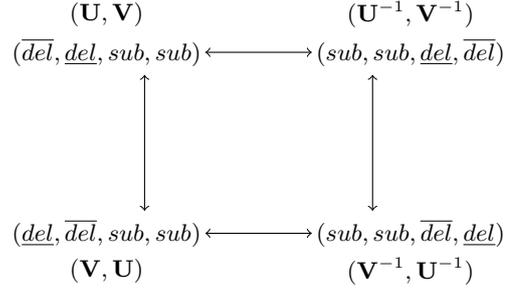
\begin{figure}[htbp]
\centering
\begin{tikzpicture}

\node at (-2,1.2) {\small $(\overline{del},\underline{del},sub,sub)$};
\node at (-2,1.7) {\small $(\mathbf{U},\mathbf{V})$};
\node at (-2,-1.2) {\small $(\underline{del},\overline{del},sub,sub)$};
\node at (-2,-1.7) {\small $(\mathbf{V},\mathbf{U})$};
\node at (2,1.2) {\small $(sub,sub,\underline{del},\overline{del})$};
\node at (2,1.7) {\small $(\mathbf{U}^{-1},\mathbf{V}^{-1})$};
\node at (2,-1.2) {\small $(sub,sub,\overline{del},\underline{del})$};
\node at (2,-1.7) {\small $(\mathbf{V}^{-1},\mathbf{U}^{-1})$};

\draw[<->] (-1.5,0.9)--(-1.5,-0.9);
\draw[<->] (1.5,0.9)--(1.5,-0.9);
\draw[<->] (-0.7,1.2)--(0.7,1.2);
\draw[<->] (-0.7,-1.2)--(0.7,-1.2);

\end{tikzpicture}
\caption{The equivalence between 
$(\overline{del},\underline{del},sub,sub)$, $(\underline{del},\overline{del},sub,sub)$, 
$(sub,sub,\underline{del},\overline{del})$, and $(sub,sub,\overline{del},\underline{del})$.}
\label{M.f6.1}
\end{figure}

As shown in  Fig. \ref{M.f6.1}, by 
$\sigma(\mathbf{F}(\mathbf{U})-\mathbf{F}(\mathbf{V}))=\sigma(\mathbf{F}(\mathbf{V})-\mathbf{F}(\mathbf{U}))$ 
and Lemma \ref{M.l6.1}, the four types of $(\mathbf{U},\mathbf{V})$, 
$(\overline{del},\underline{del},sub,sub)$, $(\underline{del},\overline{del},sub,sub)$, 
$(sub,sub,\underline{del},\overline{del})$, and $(sub,sub,\overline{del},\underline{del})$, 
are considered to be equivalent. 
Note that after this transformation, the pairwise distances of errors are at least $4$, 
which still satisfy the requirement of $(1,1)$-del/sub good pair. Analogously, 
$(\overline{del},sub,\underline{del},sub)$, $(\underline{del},sub,\overline{del},sub)$, 
$(sub,\underline{del},sub,\overline{del})$, and $(sub,\overline{del},sub,\underline{del})$ are equivalent. 
$(\overline{del},sub,sub,\underline{del})$ and $(\underline{del},sub,sub,\overline{del})$ are equivalent. 
$(sub,\overline{del},\underline{del},sub)$ and $(sub,\underline{del},\overline{del},sub)$ are equivalent. 
We divide our discussion into four non-equivalent subcases by the type of $(\mathbf{U},\mathbf{V})$. 




\subsection{The Type of $(\mathbf{U},\mathbf{V})$ is 
$(\overline{del},\underline{del},sub,sub)$}\label{sec6.1}

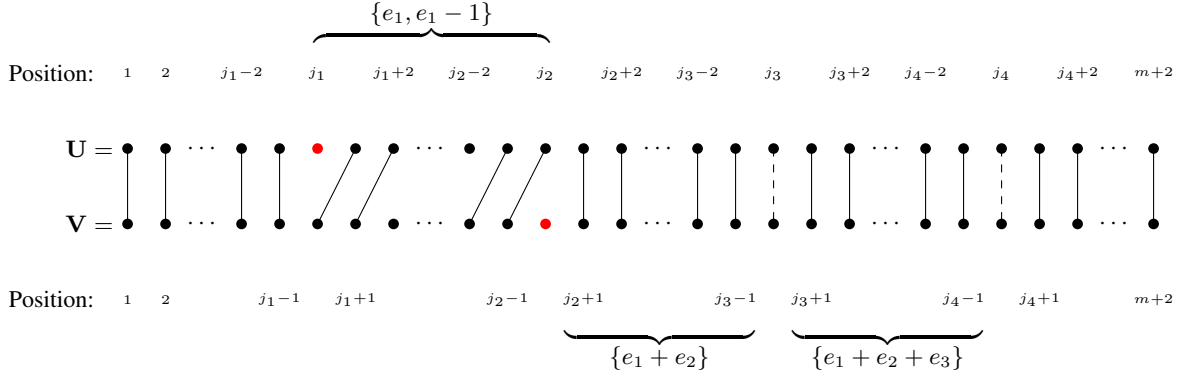
\begin{figure*}[htbp]
\centering
\begin{tikzpicture}

\node at (-0.5,1.5) {\small Position:};

\node at (0.5,1.5) {\tiny $1$};

\node at (1,1.5) {\tiny $2$};

\node at (2,1.5) {\tiny $j_{1}$$-$$2$};

\node at (3,1.5) {\tiny $j_{1}$};

\node at (4,1.5) {\tiny $j_{1}$$+$$2$};

\node at (5,1.5) {\tiny $j_{2}$$-$$2$};

\node at (6,1.5) {\tiny $j_{2}$};

\node at (7,1.5) {\tiny $j_{2}$$+$$2$};

\node at (8,1.5) {\tiny $j_{3}$$-$$2$};

\node at (9,1.5) {\tiny $j_{3}$};

\node at (10,1.5) {\tiny $j_{3}$$+$$2$};

\node at (11,1.5) {\tiny $j_{4}$$-$$2$};

\node at (12,1.5) {\tiny $j_{4}$};

\node at (13,1.5) {\tiny $j_{4}$$+$$2$};

\node at (14,1.5) {\tiny $m$$+$$2$};

\node at (-0.5,-1.5) {\small Position:};

\node at (0.5,-1.5) {\tiny $1$};

\node at (1,-1.5) {\tiny $2$};

\node at (2.5,-1.5) {\tiny $j_{1}$$-$$1$};

\node at (3.5,-1.5) {\tiny $j_{1}$$+$$1$};

\node at (5.5,-1.5) {\tiny $j_{2}$$-$$1$};

\node at (6.5,-1.5) {\tiny $j_{2}$$+$$1$};

\node at (8.5,-1.5) {\tiny $j_{3}$$-$$1$};

\node at (9.5,-1.5) {\tiny $j_{3}$$+$$1$};

\node at (11.5,-1.5) {\tiny $j_{4}$$-$$1$};

\node at (12.5,-1.5) {\tiny $j_{4}$$+$$1$};

\node at (14,-1.5) {\tiny $m$$+$$2$};

\node[rotate=180] at (4.5,2) {$\underbrace{\hspace{3.1cm}}$};
\node[above] at (4.5,2) {\small $\{ e_{1},e_{1}-1 \}$};

\node at (7.5,-2) {$\underbrace{\hspace{2.5cm}}$};
\node[below] at (7.5,-2) {\small $\{ e_{1}+e_{2} \}$};

\node at (10.5,-2) {$\underbrace{\hspace{2.5cm}}$};
\node[below] at (10.5,-2) {\small $\{ e_{1}+e_{2}+e_{3} \}$};

\node at (0,0.5) {\small$\mathbf{U}=$};

\fill (0.5,0.5) circle (2pt);

\fill (1,0.5) circle (2pt);

\node at (1.5,0.5) {\small $\cdots$};

\fill (2,0.5) circle (2pt);

\fill (2.5,0.5) circle (2pt);

\fill[red] (3,0.5) circle (2pt);

\fill (3.5,0.5) circle (2pt);

\fill (4,0.5) circle (2pt);

\node at (4.5,0.5) {\small $\cdots$};

\fill (5,0.5) circle (2pt);

\fill (5.5,0.5) circle (2pt);

\fill (6,0.5) circle (2pt);

\fill (6.5,0.5) circle (2pt);

\fill (7,0.5) circle (2pt);

\node at (7.5,0.5) {\small $\cdots$};

\fill (8,0.5) circle (2pt);

\fill (8.5,0.5) circle (2pt);

\fill (9,0.5) circle (2pt);

\fill (9.5,0.5) circle (2pt);

\fill (10,0.5) circle (2pt);

\node at (10.5,0.5) {\small $\cdots$};

\fill (11,0.5) circle (2pt);

\fill (11.5,0.5) circle (2pt);

\fill (12,0.5) circle (2pt);

\fill (12.5,0.5) circle (2pt);

\fill (13,0.5) circle (2pt);

\node at (13.5,0.5) {\small $\cdots$};

\fill (14,0.5) circle (2pt);

\node at (0,-0.5) {\small$\mathbf{V}=$};

\fill (0.5,-0.5) circle (2pt);

\fill (1,-0.5) circle (2pt);

\node at (1.5,-0.5) {\small $\cdots$};

\fill (2,-0.5) circle (2pt);

\fill (2.5,-0.5) circle (2pt);

\fill (3,-0.5) circle (2pt);

\fill (3.5,-0.5) circle (2pt);

\fill (4,-0.5) circle (2pt);

\node at (4.5,-0.5) {\small $\cdots$};

\fill (5,-0.5) circle (2pt);

\fill (5.5,-0.5) circle (2pt);

\fill[red] (6,-0.5) circle (2pt);

\fill (6.5,-0.5) circle (2pt);

\fill (7,-0.5) circle (2pt);

\node at (7.5,-0.5) {\small $\cdots$};

\fill (8,-0.5) circle (2pt);

\fill (8.5,-0.5) circle (2pt);

\fill (9,-0.5) circle (2pt);

\fill (9.5,-0.5) circle (2pt);

\fill (10,-0.5) circle (2pt);

\node at (10.5,-0.5) {\small $\cdots$};

\fill (11,-0.5) circle (2pt);

\fill (11.5,-0.5) circle (2pt);

\fill (12,-0.5) circle (2pt);

\fill (12.5,-0.5) circle (2pt);

\fill (13,-0.5) circle (2pt);

\node at (13.5,-0.5) {\small $\cdots$};

\fill (14,-0.5) circle (2pt);

\draw (0.5,0.5)--(0.5,-0.5);

\draw (1,0.5)--(1,-0.5);

\draw (2,0.5)--(2,-0.5);

\draw (2.5,0.5)--(2.5,-0.5);

\draw (3.5,0.5)--(3,-0.5);

\draw (4,0.5)--(3.5,-0.5);

\draw (5.5,0.5)--(5,-0.5);

\draw (6,0.5)--(5.5,-0.5);

\draw (6.5,0.5)--(6.5,-0.5);

\draw (7,0.5)--(7,-0.5);

\draw (8,0.5)--(8,-0.5);

\draw (8.5,0.5)--(8.5,-0.5);

\draw[dashed] (9,0.5)--(9,-0.5);

\draw (9.5,0.5)--(9.5,-0.5);

\draw (10,0.5)--(10,-0.5);

\draw (11,0.5)--(11,-0.5);

\draw (11.5,0.5)--(11.5,-0.5);

\draw[dashed] (12,0.5)--(12,-0.5);

\draw (12.5,0.5)--(12.5,-0.5);

\draw (13,0.5)--(13,-0.5);

\draw (14,0.5)--(14,-0.5);

\end{tikzpicture}
\caption{The type of $(\mathbf{U},\mathbf{V})$ is $(\overline{del},\underline{del},sub,sub)$.}
\label{M.f6.a}
\end{figure*}

Assume the type value of $(\mathbf{U},\mathbf{V})$ is $(\overline{e_{1}},\underline{e_{2}},e_{3},e_{4})$. 
That is to say, deletion of type value $\overline{e_{1}}$, deletion of type value $\underline{e_{2}}$, 
substitution of type value $e_{3}$, and substitution of type value $e_{4}$ occur at 
$U_{j_{1}}$, $V_{j_{2}}$, $U_{j_{3}}$, and $U_{j_{4}}$, respectively. 
Referring to Fig. \ref{M.f6.a}, $e_{1}=f(\mathbf{U}_{[j_{1}-1,j_{1}+1]})-f(\mathbf{V}_{[j_{1}-1,j_{1}]})$, 
$e_{2}=f(\mathbf{U}_{[j_{2},j_{2}+1]})-f(\mathbf{V}_{[j_{2}-1,j_{2}+1]})$, 
$e_{3}=f(\mathbf{U}_{[j_{3}-1,j_{3}+1]})-f(\mathbf{V}_{[j_{3}-1,j_{3}+1]})$, 
$e_{4}=f(\mathbf{U}_{[j_{4}-1,j_{4}+1]})-f(\mathbf{V}_{[j_{4}-1,j_{4}+1]})$. 
We discuss $f_{i}$ for $i\in [m+2]$ in this subcase. 

\begin{itemize}
\item For $i\in [1,j_{1}-1]$, $f_{i}=0$.

\item For $i=j_{1}$,
\begin{equation*}
\begin{split}
f_{i}=&f(\mathbf{U}_{[j_{1}-1,j_{1}]})-f(\mathbf{V}_{[j_{1}-1,j_{1}]})
\\
=&f(\mathbf{U}_{[j_{1}-1,j_{1}+1]})-f(\mathbf{V}_{[j_{1}-1,j_{1}]})-f(\mathbf{U}_{[j_{1},j_{1}+1]})
\\
=&e_{1}-f(\mathbf{U}_{[j_{1},j_{1}+1]})
\\
\in&\{ e_{1},e_{1}-1 \}.
\end{split}
\end{equation*}

\item For $i\in [j_{1}+1,j_{2}]$, 
\begin{equation*}
\begin{split}
f_{i}=e_{1}-f(\mathbf{V}_{[i-1,i]})\in\{ e_{1},e_{1}-1 \}.
\end{split}
\end{equation*}

\item For $i\in [j_{2}+1,j_{3}-1]$, $f_{i}=e_{1}+e_{2}$.

\item For $i=j_{3}$, $f_{i}$ necessitates individual discussion. However, we only need to use 
$f_{i}\in\{ e_{1}+e_{2}+1,e_{1}+e_{2},e_{1}+e_{2}-1 \}$ in this subcase.

\item For $i\in [j_{3}+1,j_{4}-1]$, $f_{i}=e_{1}+e_{2}+e_{3}$.

\item For $i=j_{4}$, $f_{i}$ necessitates individual discussion. However, we only need to use 
$f_{i}\in\{ e_{1}+e_{2}+e_{3}+1,e_{1}+e_{2}+e_{3},e_{1}+e_{2}+e_{3}-1 \}$ in this subcase.

\item For $i\in [j_{4}+1,m+2]$, $f_{i}=0$.

\end{itemize}

Noting that $e_{1},e_{2},e_{3}\in\{ 2,0,-2 \}$, we obtain 
\begin{equation*}
\begin{split}
&\sigma(\mathbf{F}(\mathbf{X})-\mathbf{F}(\mathbf{Y}))
\\
\leq&\sigma(\mathbf{F}(\mathbf{U})-\mathbf{F}(\mathbf{V}))
\\
\leq&\sigma((f_{1},\cdots,f_{j_{2}}))+\sigma((f_{j_{2}+1},\cdots,f_{j_{3}}))
\\
&+\sigma((f_{j_{3}+1},\cdots,f_{m+2}))
\\
=&3,
\end{split}
\end{equation*}
which implies $\mathbf{x}=\mathbf{y}$. 




\subsection{The Type of $(\mathbf{U},\mathbf{V})$ is 
$(\overline{del},sub,\underline{del},sub)$}\label{sec6.2}

\begin{figure*}[htbp]
\centering
\begin{tikzpicture}

\node at (-0.5,1.5) {\small Position:};

\node at (0.5,1.5) {\tiny $1$};

\node at (1,1.5) {\tiny $2$};

\node at (2,1.5) {\tiny $j_{1}$$-$$2$};

\node at (3,1.5) {\tiny $j_{1}$};

\node at (4,1.5) {\tiny $j_{1}$$+$$2$};

\node at (5.5,1.5) {\tiny $j_{2}$$-$$1$};

\node at (6.5,1.5) {\tiny $j_{2}$$+$$1$};

\node at (8.5,1.5) {\tiny $j_{3}$$-$$1$};

\node at (9.5,1.5) {\tiny $j_{3}$$+$$1$};

\node at (11.5,1.5) {\tiny $j_{4}$$-$$1$};

\node at (12.5,1.5) {\tiny $j_{4}$$+$$1$};

\node at (14,1.5) {\tiny $m$$+$$2$};

\node at (-0.5,-1.5) {\small Position:};

\node at (0.5,-1.5) {\tiny $1$};

\node at (1,-1.5) {\tiny $2$};

\node at (2.5,-1.5) {\tiny $j_{1}$$-$$1$};

\node at (3.5,-1.5) {\tiny $j_{1}$$+$$1$};

\node at (5,-1.5) {\tiny $j_{2}$$-$$2$};

\node at (6,-1.5) {\tiny $j_{2}$};

\node at (7,-1.5) {\tiny $j_{2}$$+$$2$};

\node at (8,-1.5) {\tiny $j_{3}$$-$$2$};

\node at (9,-1.5) {\tiny $j_{3}$};

\node at (10,-1.5) {\tiny $j_{3}$$+$$2$};

\node at (11,-1.5) {\tiny $j_{4}$$-$$2$};

\node at (12,-1.5) {\tiny $j_{4}$};

\node at (13,-1.5) {\tiny $j_{4}$$+$$2$};

\node at (14,-1.5) {\tiny $m$$+$$2$};

\node[rotate=180] at (4.3,2) {$\underbrace{\hspace{2.8cm}}$};
\node[above] at (4.3,2) {\small $\{ e_{1},e_{1}-1 \}$};

\node at (7.5,-2) {$\underbrace{\hspace{3.2cm}}$};
\node[below] at (7.5,-2) {\small $\{ e_{1}+e_{2},e_{1}+e_{2}-1 \}$};

\node[rotate=180] at (10.5,2) {$\underbrace{\hspace{2.5cm}}$};
\node[above] at (10.5,2) {\small $\{ e_{1}+e_{2}+e_{3} \}$};

\node at (0,0.5) {\small$\mathbf{U}=$};

\fill (0.5,0.5) circle (2pt);

\fill (1,0.5) circle (2pt);

\node at (1.5,0.5) {\small $\cdots$};

\fill (2,0.5) circle (2pt);

\fill (2.5,0.5) circle (2pt);

\fill[red] (3,0.5) circle (2pt);

\fill (3.5,0.5) circle (2pt);

\fill (4,0.5) circle (2pt);

\node at (4.5,0.5) {\small $\cdots$};

\fill (5,0.5) circle (2pt);

\fill (5.5,0.5) circle (2pt);

\fill (6,0.5) circle (2pt);

\fill (6.5,0.5) circle (2pt);

\fill (7,0.5) circle (2pt);

\node at (7.5,0.5) {\small $\cdots$};

\fill (8,0.5) circle (2pt);

\fill (8.5,0.5) circle (2pt);

\fill (9,0.5) circle (2pt);

\fill (9.5,0.5) circle (2pt);

\fill (10,0.5) circle (2pt);

\node at (10.5,0.5) {\small $\cdots$};

\fill (11,0.5) circle (2pt);

\fill (11.5,0.5) circle (2pt);

\fill (12,0.5) circle (2pt);

\fill (12.5,0.5) circle (2pt);

\fill (13,0.5) circle (2pt);

\node at (13.5,0.5) {\small $\cdots$};

\fill (14,0.5) circle (2pt);

\node at (0,-0.5) {\small$\mathbf{V}=$};

\fill (0.5,-0.5) circle (2pt);

\fill (1,-0.5) circle (2pt);

\node at (1.5,-0.5) {\small $\cdots$};

\fill (2,-0.5) circle (2pt);

\fill (2.5,-0.5) circle (2pt);

\fill (3,-0.5) circle (2pt);

\fill (3.5,-0.5) circle (2pt);

\fill (4,-0.5) circle (2pt);

\node at (4.5,-0.5) {\small $\cdots$};

\fill (5,-0.5) circle (2pt);

\fill (5.5,-0.5) circle (2pt);

\fill (6,-0.5) circle (2pt);

\fill (6.5,-0.5) circle (2pt);

\fill (7,-0.5) circle (2pt);

\node at (7.5,-0.5) {\small $\cdots$};

\fill (8,-0.5) circle (2pt);

\fill (8.5,-0.5) circle (2pt);

\fill[red] (9,-0.5) circle (2pt);

\fill (9.5,-0.5) circle (2pt);

\fill (10,-0.5) circle (2pt);

\node at (10.5,-0.5) {\small $\cdots$};

\fill (11,-0.5) circle (2pt);

\fill (11.5,-0.5) circle (2pt);

\fill (12,-0.5) circle (2pt);

\fill (12.5,-0.5) circle (2pt);

\fill (13,-0.5) circle (2pt);

\node at (13.5,-0.5) {\small $\cdots$};

\fill (14,-0.5) circle (2pt);

\draw (0.5,0.5)--(0.5,-0.5);

\draw (1,0.5)--(1,-0.5);

\draw (2,0.5)--(2,-0.5);

\draw (2.5,0.5)--(2.5,-0.5);

\draw (3.5,0.5)--(3,-0.5);

\draw (4,0.5)--(3.5,-0.5);

\draw (5.5,0.5)--(5,-0.5);

\draw[dashed] (6,0.5)--(5.5,-0.5);

\draw (6.5,0.5)--(6,-0.5);

\draw (7,0.5)--(6.5,-0.5);

\draw (8.5,0.5)--(8,-0.5);

\draw (9,0.5)--(8.5,-0.5);

\draw (9.5,0.5)--(9.5,-0.5);

\draw (10,0.5)--(10,-0.5);

\draw (11,0.5)--(11,-0.5);

\draw (11.5,0.5)--(11.5,-0.5);

\draw[dashed] (12,0.5)--(12,-0.5);

\draw (12.5,0.5)--(12.5,-0.5);

\draw (13,0.5)--(13,-0.5);

\draw (14,0.5)--(14,-0.5);

\end{tikzpicture}
\caption{The type of $(\mathbf{U},\mathbf{V})$ is $(\overline{del},sub,\underline{del},sub)$.}
\label{M.f6.b}
\end{figure*}
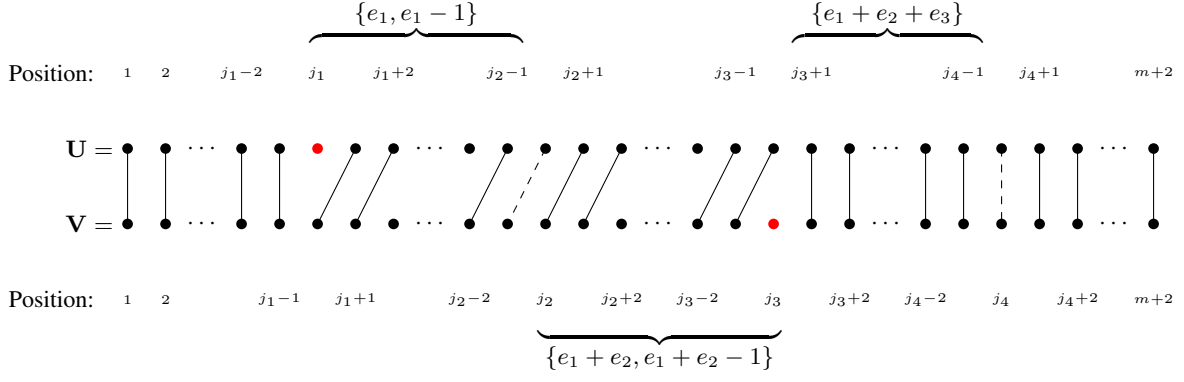

Assume the type value of $(\mathbf{U},\mathbf{V})$ is $(\overline{e_{1}},e_{2},\underline{e_{3}},e_{4})$. 
That is to say, deletion of type value $\overline{e_{1}}$, substitution of type value $e_{2}$, 
deletion of type value $\underline{e_{3}}$, and substitution of type value $e_{4}$ occur at 
$U_{j_{1}}$, $U_{j_{2}}$, $V_{j_{3}}$, and $U_{j_{4}}$, respectively. 
Referring to Fig. \ref{M.f6.b}, $e_{1}=f(\mathbf{U}_{[j_{1}-1,j_{1}+1]})-f(\mathbf{V}_{[j_{1}-1,j_{1}]})$, 
$e_{2}=f(\mathbf{U}_{[j_{2}-1,j_{2}+1]})-f(\mathbf{V}_{[j_{2}-2,j_{2}]})$, 
$e_{3}=f(\mathbf{U}_{[j_{3},j_{3}+1]})-f(\mathbf{V}_{[j_{3}-1,j_{3}+1]})$, 
$e_{4}=f(\mathbf{U}_{[j_{4}-1,j_{4}+1]})-f(\mathbf{V}_{[j_{4}-1,j_{4}+1]})$. 
We discuss $f_{i}$ for $i\in [m+2]$ in this subcase. 

\begin{itemize}
\item For $i\in [1,j_{1}-1]$, $f_{i}=0$.

\item For $i=j_{1}$,
\begin{equation*}
\begin{split}
f_{i}=&f(\mathbf{U}_{[j_{1}-1,j_{1}]})-f(\mathbf{V}_{[j_{1}-1,j_{1}]})
\\
=&f(\mathbf{U}_{[j_{1}-1,j_{1}+1]})-f(\mathbf{V}_{[j_{1}-1,j_{1}]})-f(\mathbf{U}_{[j_{1},j_{1}+1]})
\\
=&e_{1}-f(\mathbf{U}_{[j_{1},j_{1}+1]})
\\
\in&\{ e_{1},e_{1}-1 \}.
\end{split}
\end{equation*}

\item For $i\in [j_{1}+1,j_{2}-1]$, 
\begin{equation*}
\begin{split}
f_{i}=e_{1}-f(\mathbf{V}_{[i-1,i]})\in\{ e_{1},e_{1}-1 \}.
\end{split}
\end{equation*}

\item For $i=j_{2}$,
\begin{equation*}
\begin{split}
f_{i}=&e_{1}+f(\mathbf{U}_{[j_{2}-1,j_{2}]})-f(\mathbf{V}_{[j_{2}-2,j_{2}]})
\\
=&e_{1}+f(\mathbf{U}_{[j_{2}-1,j_{2}+1]})-f(\mathbf{V}_{[j_{2}-2,j_{2}]})
\\
&-f(\mathbf{U}_{[j_{2},j_{2}+1]})
\\
=&e_{1}+e_{2}-f(\mathbf{U}_{[j_{2},j_{2}+1]})
\\
\in&\{ e_{1}+e_{2},e_{1}+e_{2}-1 \}.
\end{split}
\end{equation*}

\item For $i\in [j_{2}+1,j_{3}]$, 
\begin{equation*}
\begin{split}
f_{i}=e_{1}+e_{2}-f(\mathbf{V}_{[i-1,i]})
\in\{ e_{1}+e_{2},e_{1}+e_{2}-1 \}.
\end{split}
\end{equation*}

\item For $i\in [j_{3}+1,j_{4}-1]$, $f_{i}=e_{1}+e_{2}+e_{3}$.

\item For $i=j_{4}$, $f_{i}$ necessitates individual discussion. However, we only need to use 
$f_{i}\in\{ e_{1}+e_{2}+e_{3}+1,e_{1}+e_{2}+e_{3},e_{1}+e_{2}+e_{3}-1 \}$ in this subcase.

\item For $i\in [j_{4}+1,m+2]$, $f_{i}=0$.

\end{itemize}

Noting that $e_{1},e_{2},e_{3}\in\{ 2,0,-2 \}$, we obtain 
\begin{equation*}
\begin{split}
&\sigma(\mathbf{F}(\mathbf{X})-\mathbf{F}(\mathbf{Y}))
\\
\leq&\sigma(\mathbf{F}(\mathbf{U})-\mathbf{F}(\mathbf{V}))
\\
\leq&\sigma((f_{1},\cdots,f_{j_{2}-1}))+\sigma((f_{j_{2}},\cdots,f_{j_{3}}))
\\
&+\sigma((f_{j_{3}+1},\cdots,f_{m+2}))
\\
=&3,
\end{split}
\end{equation*}
which implies $\mathbf{x}=\mathbf{y}$. 




\subsection{The Type of $(\mathbf{U},\mathbf{V})$ is 
$(\overline{del},sub,sub,\underline{del})$}\label{sec6.3}

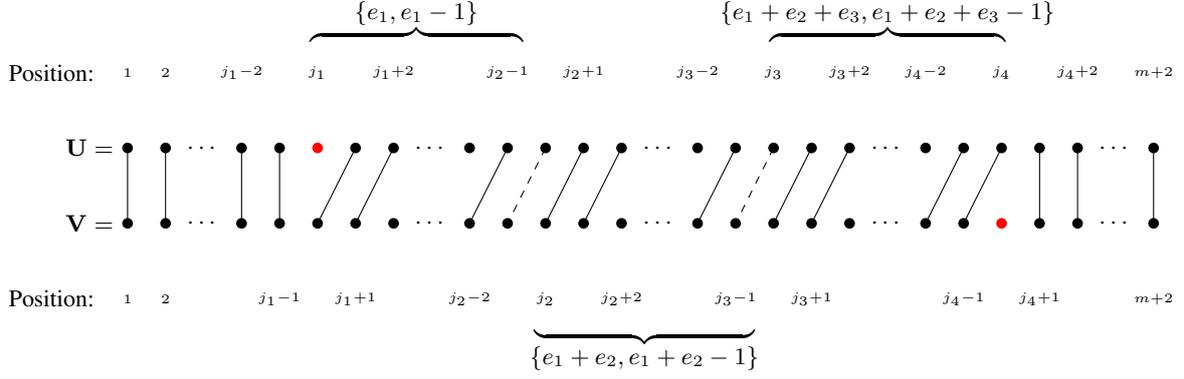
\begin{figure*}[htbp]
\centering
\begin{tikzpicture}

\node at (-0.5,1.5) {\small Position:};

\node at (0.5,1.5) {\tiny $1$};

\node at (1,1.5) {\tiny $2$};

\node at (2,1.5) {\tiny $j_{1}$$-$$2$};

\node at (3,1.5) {\tiny $j_{1}$};

\node at (4,1.5) {\tiny $j_{1}$$+$$2$};

\node at (5.5,1.5) {\tiny $j_{2}$$-$$1$};

\node at (6.5,1.5) {\tiny $j_{2}$$+$$1$};

\node at (8,1.5) {\tiny $j_{3}$$-$$2$};

\node at (9,1.5) {\tiny $j_{3}$};

\node at (10,1.5) {\tiny $j_{3}$$+$$2$};

\node at (11,1.5) {\tiny $j_{4}$$-$$2$};

\node at (12,1.5) {\tiny $j_{4}$};

\node at (13,1.5) {\tiny $j_{4}$$+$$2$};

\node at (14,1.5) {\tiny $m$$+$$2$};

\node at (-0.5,-1.5) {\small Position:};

\node at (0.5,-1.5) {\tiny $1$};

\node at (1,-1.5) {\tiny $2$};

\node at (2.5,-1.5) {\tiny $j_{1}$$-$$1$};

\node at (3.5,-1.5) {\tiny $j_{1}$$+$$1$};

\node at (5,-1.5) {\tiny $j_{2}$$-$$2$};

\node at (6,-1.5) {\tiny $j_{2}$};

\node at (7,-1.5) {\tiny $j_{2}$$+$$2$};

\node at (8.5,-1.5) {\tiny $j_{3}$$-$$1$};

\node at (9.5,-1.5) {\tiny $j_{3}$$+$$1$};

\node at (11.5,-1.5) {\tiny $j_{4}$$-$$1$};

\node at (12.5,-1.5) {\tiny $j_{4}$$+$$1$};

\node at (14,-1.5) {\tiny $m$$+$$2$};

\node[rotate=180] at (4.3,2) {$\underbrace{\hspace{2.8cm}}$};
\node[above] at (4.3,2) {\small $\{ e_{1},e_{1}-1 \}$};

\node at (7.3,-2) {$\underbrace{\hspace{2.9cm}}$};
\node[below] at (7.3,-2) {\small $\{ e_{1}+e_{2},e_{1}+e_{2}-1 \}$};

\node[rotate=180] at (10.5,2) {$\underbrace{\hspace{3.1cm}}$};
\node[above] at (10.5,2) {\small $\{ e_{1}+e_{2}+e_{3},e_{1}+e_{2}+e_{3}-1 \}$};

\node at (0,0.5) {\small$\mathbf{U}=$};

\fill (0.5,0.5) circle (2pt);

\fill (1,0.5) circle (2pt);

\node at (1.5,0.5) {\small $\cdots$};

\fill (2,0.5) circle (2pt);

\fill (2.5,0.5) circle (2pt);

\fill[red] (3,0.5) circle (2pt);

\fill (3.5,0.5) circle (2pt);

\fill (4,0.5) circle (2pt);

\node at (4.5,0.5) {\small $\cdots$};

\fill (5,0.5) circle (2pt);

\fill (5.5,0.5) circle (2pt);

\fill (6,0.5) circle (2pt);

\fill (6.5,0.5) circle (2pt);

\fill (7,0.5) circle (2pt);

\node at (7.5,0.5) {\small $\cdots$};

\fill (8,0.5) circle (2pt);

\fill (8.5,0.5) circle (2pt);

\fill (9,0.5) circle (2pt);

\fill (9.5,0.5) circle (2pt);

\fill (10,0.5) circle (2pt);

\node at (10.5,0.5) {\small $\cdots$};

\fill (11,0.5) circle (2pt);

\fill (11.5,0.5) circle (2pt);

\fill (12,0.5) circle (2pt);

\fill (12.5,0.5) circle (2pt);

\fill (13,0.5) circle (2pt);

\node at (13.5,0.5) {\small $\cdots$};

\fill (14,0.5) circle (2pt);

\node at (0,-0.5) {\small$\mathbf{V}=$};

\fill (0.5,-0.5) circle (2pt);

\fill (1,-0.5) circle (2pt);

\node at (1.5,-0.5) {\small $\cdots$};

\fill (2,-0.5) circle (2pt);

\fill (2.5,-0.5) circle (2pt);

\fill (3,-0.5) circle (2pt);

\fill (3.5,-0.5) circle (2pt);

\fill (4,-0.5) circle (2pt);

\node at (4.5,-0.5) {\small $\cdots$};

\fill (5,-0.5) circle (2pt);

\fill (5.5,-0.5) circle (2pt);

\fill (6,-0.5) circle (2pt);

\fill (6.5,-0.5) circle (2pt);

\fill (7,-0.5) circle (2pt);

\node at (7.5,-0.5) {\small $\cdots$};

\fill (8,-0.5) circle (2pt);

\fill (8.5,-0.5) circle (2pt);

\fill (9,-0.5) circle (2pt);

\fill (9.5,-0.5) circle (2pt);

\fill (10,-0.5) circle (2pt);

\node at (10.5,-0.5) {\small $\cdots$};

\fill (11,-0.5) circle (2pt);

\fill (11.5,-0.5) circle (2pt);

\fill[red] (12,-0.5) circle (2pt);

\fill (12.5,-0.5) circle (2pt);

\fill (13,-0.5) circle (2pt);

\node at (13.5,-0.5) {\small $\cdots$};

\fill (14,-0.5) circle (2pt);

\draw (0.5,0.5)--(0.5,-0.5);

\draw (1,0.5)--(1,-0.5);

\draw (2,0.5)--(2,-0.5);

\draw (2.5,0.5)--(2.5,-0.5);

\draw (3.5,0.5)--(3,-0.5);

\draw (4,0.5)--(3.5,-0.5);

\draw (5.5,0.5)--(5,-0.5);

\draw[dashed] (6,0.5)--(5.5,-0.5);

\draw (6.5,0.5)--(6,-0.5);

\draw (7,0.5)--(6.5,-0.5);

\draw (8.5,0.5)--(8,-0.5);

\draw[dashed] (9,0.5)--(8.5,-0.5);

\draw (9.5,0.5)--(9,-0.5);

\draw (10,0.5)--(9.5,-0.5);

\draw (11.5,0.5)--(11,-0.5);

\draw (12,0.5)--(11.5,-0.5);

\draw (12.5,0.5)--(12.5,-0.5);

\draw (13,0.5)--(13,-0.5);

\draw (14,0.5)--(14,-0.5);

\end{tikzpicture}
\caption{The type of $(\mathbf{U},\mathbf{V})$ is $(\overline{del},sub,sub,\underline{del})$.}
\label{M.f6.c}
\end{figure*}

Assume the type value of $(\mathbf{U},\mathbf{V})$ is $(\overline{e_{1}},e_{2},e_{3},\underline{e_{4}})$. 
That is to say, deletion of type value $\overline{e_{1}}$, substitution of type value $e_{2}$, 
substitution of type value $e_{3}$, and deletion of type value $\underline{e_{4}}$ occur at 
$U_{j_{1}}$, $U_{j_{2}}$, $U_{j_{3}}$, and $V_{j_{4}}$, respectively. 
Referring to Fig. \ref{M.f6.c}, $e_{1}=f(\mathbf{U}_{[j_{1}-1,j_{1}+1]})-f(\mathbf{V}_{[j_{1}-1,j_{1}]})$, 
$e_{2}=f(\mathbf{U}_{[j_{2}-1,j_{2}+1]})-f(\mathbf{V}_{[j_{2}-2,j_{2}]})$, 
$e_{3}=f(\mathbf{U}_{[j_{3}-1,j_{3}+1]})-f(\mathbf{V}_{[j_{3}-2,j_{3}]})$, 
$e_{4}=f(\mathbf{U}_{[j_{4},j_{4}+1]})-f(\mathbf{V}_{[j_{4}-1,j_{4}+1]})$. 
We discuss $f_{i}$ for $i\in [m+2]$ in this subcase. 

\begin{itemize}
\item For $i\in [1,j_{1}-1]$, $f_{i}=0$.

\item For $i=j_{1}$,
\begin{equation*}
\begin{split}
f_{i}=&f(\mathbf{U}_{[j_{1}-1,j_{1}]})-f(\mathbf{V}_{[j_{1}-1,j_{1}]})
\\
=&f(\mathbf{U}_{[j_{1}-1,j_{1}+1]})-f(\mathbf{V}_{[j_{1}-1,j_{1}]})-f(\mathbf{U}_{[j_{1},j_{1}+1]})
\\
=&e_{1}-f(\mathbf{U}_{[j_{1},j_{1}+1]})
\\
\in&\{ e_{1},e_{1}-1 \}.
\end{split}
\end{equation*}

\item For $i\in [j_{1}+1,j_{2}-1]$, 
\begin{equation*}
\begin{split}
f_{i}=e_{1}-f(\mathbf{V}_{[i-1,i]})\in\{ e_{1},e_{1}-1 \}.
\end{split}
\end{equation*}

\item For $i=j_{2}$,
\begin{equation*}
\begin{split}
f_{i}=&e_{1}+f(\mathbf{U}_{[j_{2}-1,j_{2}]})-f(\mathbf{V}_{[j_{2}-2,j_{2}]})
\\
=&e_{1}+f(\mathbf{U}_{[j_{2}-1,j_{2}+1]})-f(\mathbf{V}_{[j_{2}-2,j_{2}]})
\\
&-f(\mathbf{U}_{[j_{2},j_{2}+1]})
\\
=&e_{1}+e_{2}-f(\mathbf{U}_{[j_{2},j_{2}+1]})
\\
\in&\{ e_{1}+e_{2},e_{1}+e_{2}-1 \}.
\end{split}
\end{equation*}

\item For $i\in [j_{2}+1,j_{3}-1]$, 
\begin{equation*}
\begin{split}
f_{i}=e_{1}+e_{2}-f(\mathbf{V}_{[i-1,i]})
\in\{ e_{1}+e_{2},e_{1}+e_{2}-1 \}.
\end{split}
\end{equation*}

\item For $i=j_{3}$,
\begin{equation*}
\begin{split}
f_{i}=&e_{1}+e_{2}+f(\mathbf{U}_{[j_{3}-1,j_{3}]})-f(\mathbf{V}_{[j_{3}-2,j_{3}]})
\\
=&e_{1}+e_{2}+f(\mathbf{U}_{[j_{3}-1,j_{3}+1]})-f(\mathbf{V}_{[j_{3}-2,j_{3}]})
\\
&-f(\mathbf{U}_{[j_{3},j_{3}+1]})
\\
=&e_{1}+e_{2}+e_{3}-f(\mathbf{U}_{[j_{3},j_{3}+1]})
\\
\in&\{ e_{1}+e_{2}+e_{3},e_{1}+e_{2}+e_{3}-1 \}.
\end{split}
\end{equation*}

\item For $i\in [j_{3}+1,j_{4}]$, 
\begin{equation*}
\begin{split}
f_{i}=&e_{1}+e_{2}+e_{3}-f(\mathbf{V}_{[i-1,i]})
\\
\in&\{ e_{1}+e_{2}+e_{3},e_{1}+e_{2}+e_{3}-1 \}.
\end{split}
\end{equation*}

\item For $i\in [j_{4}+1,m+2]$, $f_{i}=0$.

\end{itemize}

Noting that $e_{1},e_{2},e_{3}\in\{ 2,0,-2 \}$, we obtain 
\begin{equation*}
\begin{split}
&\sigma(\mathbf{F}(\mathbf{X})-\mathbf{F}(\mathbf{Y}))
\\
\leq&\sigma(\mathbf{F}(\mathbf{U})-\mathbf{F}(\mathbf{V}))
\\
\leq&\sigma((f_{1},\cdots,f_{j_{2}-1}))+\sigma((f_{j_{2}},\cdots,f_{j_{3}-1}))
\\
&+\sigma((f_{j_{3}},\cdots,f_{m+2}))
\\
=&3,
\end{split}
\end{equation*}
which implies $\mathbf{x}=\mathbf{y}$. 




\subsection{The Type of $(\mathbf{U},\mathbf{V})$ is 
$(sub,\overline{del},\underline{del},sub)$}\label{sec6.4}

\begin{figure*}[htbp]
\centering
\begin{tikzpicture}

\node at (-0.5,1.5) {\small Position:};

\node at (0.5,1.5) {\tiny $1$};

\node at (1,1.5) {\tiny $2$};

\node at (2.5,1.5) {\tiny $j_{1}$$-$$1$};

\node at (3.5,1.5) {\tiny $j_{1}$$+$$1$};

\node at (5.5,1.5) {\tiny $j_{2}$$-$$1$};

\node at (6.5,1.5) {\tiny $j_{2}$$+$$1$};

\node at (8.5,1.5) {\tiny $j_{3}$$-$$1$};

\node at (9.5,1.5) {\tiny $j_{3}$$+$$1$};

\node at (11.5,1.5) {\tiny $j_{4}$$-$$1$};

\node at (12.5,1.5) {\tiny $j_{4}$$+$$1$};

\node at (14,1.5) {\tiny $m$$+$$2$};

\node at (-0.5,-1.5) {\small Position:};

\node at (0.5,-1.5) {\tiny $1$};

\node at (1,-1.5) {\tiny $2$};

\node at (2,-1.5) {\tiny $j_{1}$$-$$2$};

\node at (3,-1.5) {\tiny $j_{1}$};

\node at (4,-1.5) {\tiny $j_{1}$$+$$2$};

\node at (5,-1.5) {\tiny $j_{2}$$-$$2$};

\node at (6,-1.5) {\tiny $j_{2}$};

\node at (7,-1.5) {\tiny $j_{2}$$+$$2$};

\node at (8,-1.5) {\tiny $j_{3}$$-$$2$};

\node at (9,-1.5) {\tiny $j_{3}$};

\node at (10,-1.5) {\tiny $j_{3}$$+$$2$};

\node at (11,-1.5) {\tiny $j_{4}$$-$$2$};

\node at (12,-1.5) {\tiny $j_{4}$};

\node at (13,-1.5) {\tiny $j_{4}$$+$$2$};

\node at (14,-1.5) {\tiny $m$$+$$2$};

\node[rotate=180] at (4.5,2) {$\underbrace{\hspace{2.5cm}}$};
\node[above] at (4.5,2) {\small $\{ e_{1} \}$};

\node at (7.5,-2) {$\underbrace{\hspace{3.2cm}}$};
\node[below] at (7.5,-2) {\small $\{ e_{1}+e_{2},e_{1}+e_{2}-1 \}$};

\node[rotate=180] at (10.5,2) {$\underbrace{\hspace{2,5cm}}$};
\node[above] at (10.5,2) {\small $\{ e_{1}+e_{2}+e_{3} \}$};

\node at (0,0.5) {\small$\mathbf{U}=$};

\fill (0.5,0.5) circle (2pt);

\fill (1,0.5) circle (2pt);

\node at (1.5,0.5) {\small $\cdots$};

\fill (2,0.5) circle (2pt);

\fill (2.5,0.5) circle (2pt);

\fill (3,0.5) circle (2pt);

\fill (3.5,0.5) circle (2pt);

\fill (4,0.5) circle (2pt);

\node at (4.5,0.5) {\small $\cdots$};

\fill (5,0.5) circle (2pt);

\fill (5.5,0.5) circle (2pt);

\fill[red] (6,0.5) circle (2pt);

\fill (6.5,0.5) circle (2pt);

\fill (7,0.5) circle (2pt);

\node at (7.5,0.5) {\small $\cdots$};

\fill (8,0.5) circle (2pt);

\fill (8.5,0.5) circle (2pt);

\fill (9,0.5) circle (2pt);

\fill (9.5,0.5) circle (2pt);

\fill (10,0.5) circle (2pt);

\node at (10.5,0.5) {\small $\cdots$};

\fill (11,0.5) circle (2pt);

\fill (11.5,0.5) circle (2pt);

\fill (12,0.5) circle (2pt);

\fill (12.5,0.5) circle (2pt);

\fill (13,0.5) circle (2pt);

\node at (13.5,0.5) {\small $\cdots$};

\fill (14,0.5) circle (2pt);

\node at (0,-0.5) {\small$\mathbf{V}=$};

\fill (0.5,-0.5) circle (2pt);

\fill (1,-0.5) circle (2pt);

\node at (1.5,-0.5) {\small $\cdots$};

\fill (2,-0.5) circle (2pt);

\fill (2.5,-0.5) circle (2pt);

\fill (3,-0.5) circle (2pt);

\fill (3.5,-0.5) circle (2pt);

\fill (4,-0.5) circle (2pt);

\node at (4.5,-0.5) {\small $\cdots$};

\fill (5,-0.5) circle (2pt);

\fill (5.5,-0.5) circle (2pt);

\fill (6,-0.5) circle (2pt);

\fill (6.5,-0.5) circle (2pt);

\fill (7,-0.5) circle (2pt);

\node at (7.5,-0.5) {\small $\cdots$};

\fill (8,-0.5) circle (2pt);

\fill (8.5,-0.5) circle (2pt);

\fill[red] (9,-0.5) circle (2pt);

\fill (9.5,-0.5) circle (2pt);

\fill (10,-0.5) circle (2pt);

\node at (10.5,-0.5) {\small $\cdots$};

\fill (11,-0.5) circle (2pt);

\fill (11.5,-0.5) circle (2pt);

\fill (12,-0.5) circle (2pt);

\fill (12.5,-0.5) circle (2pt);

\fill (13,-0.5) circle (2pt);

\node at (13.5,-0.5) {\small $\cdots$};

\fill (14,-0.5) circle (2pt);

\draw (0.5,0.5)--(0.5,-0.5);

\draw (1,0.5)--(1,-0.5);

\draw (2,0.5)--(2,-0.5);

\draw (2.5,0.5)--(2.5,-0.5);

\draw[dashed] (3,0.5)--(3,-0.5);

\draw (3.5,0.5)--(3.5,-0.5);

\draw (4,0.5)--(4,-0.5);

\draw (5,0.5)--(5,-0.5);

\draw (5.5,0.5)--(5.5,-0.5);

\draw (6.5,0.5)--(6,-0.5);

\draw (6.5,0.5)--(6,-0.5);

\draw (7,0.5)--(6.5,-0.5);

\draw (8.5,0.5)--(8,-0.5);

\draw (9,0.5)--(8.5,-0.5);

\draw (9.5,0.5)--(9.5,-0.5);

\draw (10,0.5)--(10,-0.5);

\draw (11,0.5)--(11,-0.5);

\draw (11.5,0.5)--(11.5,-0.5);

\draw[dashed] (12,0.5)--(12,-0.5);

\draw (12.5,0.5)--(12.5,-0.5);

\draw (13,0.5)--(13,-0.5);

\draw (14,0.5)--(14,-0.5);

\end{tikzpicture}
\caption{The type of $(\mathbf{U},\mathbf{V})$ is $(sub,\overline{del},\underline{del},sub)$.}
\label{M.f6.d}
\end{figure*}
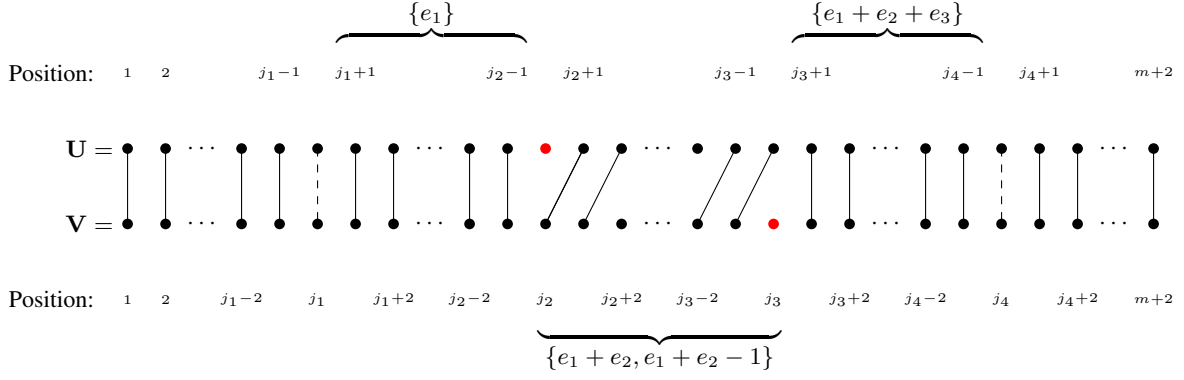

Assume the type value of $(\mathbf{U},\mathbf{V})$ is $(e_{1},\overline{e_{2}},\underline{e_{3}},e_{4})$. 
That is to say, substitution of type value $e_{1}$, deletion of type value $\overline{e_{2}}$, 
deletion of type value $\underline{e_{3}}$, and substitution of type value $e_{4}$ occur at 
$U_{j_{1}}$, $U_{j_{2}}$, $V_{j_{3}}$, and $U_{j_{4}}$, respectively. 
Referring to Fig. \ref{M.f6.d}, $e_{1}=f(\mathbf{U}_{[j_{1}-1,j_{1}+1]})-f(\mathbf{V}_{[j_{1}-1,j_{1}+1]})$, 
$e_{2}=f(\mathbf{U}_{[j_{2}-1,j_{2}+1]})-f(\mathbf{V}_{[j_{2}-1,j_{2}]})$, 
$e_{3}=f(\mathbf{U}_{[j_{3},j_{3}+1]})-f(\mathbf{V}_{[j_{3}-1,j_{3}+1]})$, 
$e_{4}=f(\mathbf{U}_{[j_{4}-1,j_{4}+1]})-f(\mathbf{V}_{[j_{4}-1,j_{4}+1]})$. 
We discuss $f_{i}$ for $i\in [m+2]$ in this subcase. 

\begin{itemize}
\item For $i\in [1,j_{1}-1]$, $f_{i}=0$.

\item For $i=j_{1}$, $f_{i}$ necessitates individual discussion. However, we only need to use 
$f_{i}\in\{ e_{1}+1,e_{1},e_{1}-1 \}$ in this subcase.

\item For $i\in [j_{1}+1,j_{2}-1]$, $f_{i}=e_{1}$.

\item For $i=j_{2}$,
\begin{equation*}
\begin{split}
f_{i}=&e_{1}+f(\mathbf{U}_{[j_{2}-1,j_{2}]})-f(\mathbf{V}_{[j_{2}-1,j_{2}]})
\\
=&e_{1}+f(\mathbf{U}_{[j_{2}-1,j_{2}+1]})-f(\mathbf{V}_{[j_{2}-1,j_{2}]})
\\
&-f(\mathbf{U}_{[j_{2},j_{2}+1]})
\\
=&e_{1}+e_{2}-f(\mathbf{U}_{[j_{2},j_{2}+1]})
\\
\in&\{ e_{1}+e_{2},e_{1}+e_{2}-1 \}.
\end{split}
\end{equation*}

\item For $i\in [j_{2}+1,j_{3}]$, 
\begin{equation*}
\begin{split}
f_{i}=e_{1}+e_{2}-f(\mathbf{V}_{[i-1,i]})\in\{ e_{1}+e_{2},e_{1}+e_{2}-1 \}.
\end{split}
\end{equation*}

\item For $i\in [j_{3}+1,j_{4}-1]$, $f_{i}=e_{1}+e_{2}+e_{3}$.

\item For $i=j_{4}$, $f_{i}$ necessitates individual discussion. However, we only need to use 
$f_{i}\in\{ e_{1}+e_{2}+e_{3}+1,e_{1}+e_{2}+e_{3},e_{1}+e_{2}+e_{3}-1 \}$ in this subcase.

\item For $i\in [j_{4}+1,m+2]$, $f_{i}=0$.

\end{itemize}

Noting that $e_{1},e_{2},e_{3}\in\{ 2,0,-2 \}$, we obtain 
\begin{equation*}
\begin{split}
&\sigma(\mathbf{F}(\mathbf{X})-\mathbf{F}(\mathbf{Y}))
\\
\leq&\sigma(\mathbf{F}(\mathbf{U})-\mathbf{F}(\mathbf{V}))
\\
\leq&\sigma((f_{1},\cdots,f_{j_{2}-1}))+\sigma((f_{j_{2}},\cdots,f_{j_{3}}))
\\
&+\sigma((f_{j_{3}+1},\cdots,f_{m+2}))
\\
=&3,
\end{split}
\end{equation*}
which implies $\mathbf{x}=\mathbf{y}$. 




\section{Conclusion}\label{sec7}
The crucial aspects of this paper center around the employment of error segmentation technique, 
enabling a rigorous classification into types and type values for $(s,r)$-del/sub good pair 
$(\mathbf{U},\mathbf{V})$. Furthermore, our discussions are primarily confined to the type of 
$(\mathbf{U},\mathbf{V})$. For each type, we proceed to set its type value to conduct unified 
analysis, which greatly reduces the burden of discussions. 




{
\appendix[Proof of Lemma \ref{M.l2.1}]

The main focus of our approach is on segmentation technique. 
Assume that the sequence obtained by deleting $X_{l_{1}},\cdots,X_{l_{s}}$ and 
substituting $X_{l_{s+1}},\cdots,X_{l_{s+2r}}$ from $\mathbf{X}$ is equal to 
the sequence obtained by deleting $Y_{l_{s+2r+1}},\cdots,Y_{l_{2s+2r}}$ from $\mathbf{Y}$ 
where $l_{1},l_{2},\cdots,l_{2s+2r}\in [2,n+1]$. 
Therefore, for any $\alpha\in [n+2]\backslash \{ l_{1},\cdots,l_{s} \}$, there exists a unique 
$\beta\in [n+2]\backslash \{ l_{s+2r+1},\cdots,l_{2s+2r} \}$ such that $X_{\alpha}$ and $Y_{\beta}$ 
are matched. We claim that, if all matched $X_{\alpha}$ and $Y_{\beta}$ satisfy 
$\begin{cases}
\alpha\leq i
\\
\beta\leq j
\end{cases}$
or
$\begin{cases}
\alpha\geq i+1
\\
\beta\geq j+1
\end{cases}$ where $1\leq i,j\leq n+1$, 
then we can apply segmentation technique to separate $\mathbf{X}_{[i]}$ and $\mathbf{X}_{[i+1,n+2]}$, 
as well as $\mathbf{Y}_{[j]}$ and $\mathbf{Y}_{[j+1,n+2]}$, simultaneously. Specifically, 
we can add an appropriate non-empty sequence $\mathbf{z}$ between $X_{i}$ and $X_{i+1}$ 
to alter $\mathbf{X}$ into $\mathbf{X}'$, and between $Y_{j}$ and $Y_{j+1}$ to alter 
$\mathbf{Y}$ into $\mathbf{Y}'$. With this segmentation in place, 
$f(\mathbf{X})-f(\mathbf{Y})=f(\mathbf{X}')-f(\mathbf{Y}')$ and 
$\sigma(\mathbf{F}(\mathbf{X})-\mathbf{F}(\mathbf{Y}))\leq 
\sigma(\mathbf{F}(\mathbf{X}')-\mathbf{F}(\mathbf{Y}'))$ hold. 
Noting that all matched $X_{\alpha}$ and $Y_{\beta}$ satisfy 
$\begin{cases}
\alpha\leq i
\\
\beta\leq j
\end{cases}$
or
$\begin{cases}
\alpha\geq i+1
\\
\beta\geq j+1
\end{cases}$, the added $\mathbf{z}$s in $\mathbf{X}$ and $\mathbf{Y}$ can be mutually matched. 

\textit{Case 1: $i=j$.} We discuss the values of 
$X_{i}$, $X_{i+1}$, $Y_{i}$, and $Y_{i+1}$ to determine 
the added $\mathbf{z}$. 
If 
$\begin{cases}
X_{i}=a
\\
Y_{i}=a
\end{cases}$, 
$\begin{cases}
X_{i+1}=a
\\
Y_{i+1}=a
\end{cases}$, 
or 
$\begin{cases}
X_{i}=Y_{i+1}=a
\\
Y_{i}=X_{i+1}=b
\end{cases}$, we select $aa$ as $\mathbf{z}$. 
Otherwise, for 
$\begin{cases}
X_{i}=X_{i+1}=a
\\
Y_{i}=Y_{i+1}=b
\end{cases}$, we select $ab$ as $\mathbf{z}$. 
One can easily verify $f(\mathbf{X}'_{[k]})-f(\mathbf{Y}'_{[k]})=f(\mathbf{X}_{[k]})-f(\mathbf{Y}_{[k]})$ 
for $1\leq k\leq i$ and $f(\mathbf{X}'_{[k]})-f(\mathbf{Y}'_{[k]})
=f(\mathbf{X}_{[k-2]})-f(\mathbf{Y}_{[k-2]})$ for $i+3\leq k\leq n+4$. As a result, 
$\mathbf{F}(\mathbf{X})-\mathbf{F}(\mathbf{Y})$ is a subsequence of 
$\mathbf{F}(\mathbf{X}')-\mathbf{F}(\mathbf{Y}')$ which implies 
$\sigma(\mathbf{F}(\mathbf{X})-\mathbf{F}(\mathbf{Y}))\leq 
\sigma(\mathbf{F}(\mathbf{X}')-\mathbf{F}(\mathbf{Y}'))$. 

\begin{figure*}[htbp]
\centering
\begin{tikzpicture}

\node at (-0.5,1.5) {\small Position:};

\node at (2.5,1.5) {\tiny $i$};

\node at (3.5,1.5) {\tiny $i$$+$$1$};

\node at (5.5,1.5) {\tiny $j$};

\node at (6.5,1.5) {\tiny $j$$+$$1$};

\node at (7.5,1.5) {\tiny $j$$+$$2$};

\node at (9.5,1.5) {\tiny $2j$$-$$i$$+1$};

\node at (10.5,1.5) {\tiny $2j$$-$$i$$+2$};

\node at (-0.5,-1.5) {\small Position:};

\node at (2.5,-1.5) {\tiny $i$};

\node at (3.5,-1.5) {\tiny $i$$+$$1$};

\node at (5.5,-1.5) {\tiny $j$};

\node at (6.5,-1.5) {\tiny $j$$+$$1$};

\node at (8.5,-1.5) {\tiny $2j$$-$$i$};

\node at (9.5,-1.5) {\tiny $2j$$-$$i$$+1$};

\node at (10.5,-1.5) {\tiny $2j$$-$$i$$+2$};

\node[rotate=180] at (5,2) {$\underbrace{\hspace{3.3cm}}$};
\node[above] at (5,2) {\small The added $\mathbf{z}$ in $\mathbf{X}$};

\node at (8.1,-2) {$\underbrace{\hspace{3.6cm}}$};
\node[below] at (8.1,-2) {\small The added $\mathbf{z}$ in $\mathbf{Y}$};

\node at (0,0.5) {\small$\mathbf{X}'=$};

\node at (1.5,0.5) {\small $\cdots$};

\node at (2.5,0.5) {\small $X_{i}$};

\node[blue] at (3.5,0.5) {\small $X_{i+1}$};

\node at (4.5,0.5) {\small $\cdots$};

\node[blue] at (5.5,0.5) {\small $X_{j}$};

\node[blue] at (6.5,0.5) {\small $Y_{j}$};

\node at (7.5,0.5) {\small $X_{i+1}$};

\node at (8.5,0.5) {\small $\cdots$};

\node at (9.5,0.5) {\small $X_{j}$};

\node at (10.5,0.5) {\small $X_{j+1}$};

\node at (0,-0.5) {\small$\mathbf{Y}'=$};

\node at (1.5,-0.5) {\small $\cdots$};

\node at (2.5,-0.5) {\small $Y_{i}$};

\node at (3.5,-0.5) {\small $Y_{i+1}$};

\node at (4.5,-0.5) {\small $\cdots$};

\node at (5.5,-0.5) {\small $Y_{j}$};

\node[blue] at (6.5,-0.5) {\small $X_{i+1}$};

\node at (7.5,-0.5) {\small $\cdots$};

\node[blue] at (8.5,-0.5) {\small $X_{j}$};

\node[blue] at (9.5,-0.5) {\small $Y_{j}$};

\node at (10.5,-0.5) {\small $Y_{j+1}$};

\draw (3.5,0.3)--(6.5,-0.3);

\draw (5.5,0.3)--(8.5,-0.3);

\draw (6.5,0.3)--(9.5,-0.3);

\end{tikzpicture}
\caption{The segmentation technique.}
\label{Appendix1}
\end{figure*}
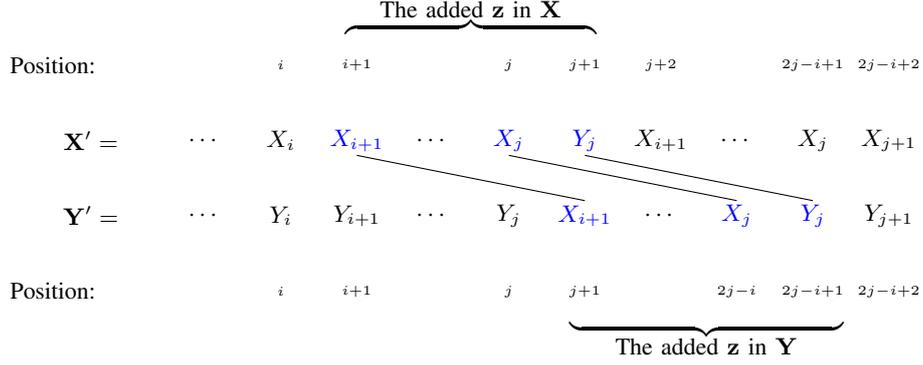

\textit{Case 2: $i<j$.} As shown in Fig. \ref{Appendix1}, we select $\mathbf{X}_{[i+1,j]}Y_{j}$ 
as $\mathbf{z}$. Clearly, $f(\mathbf{X}'_{[k]})-f(\mathbf{Y}'_{[k]})=f(\mathbf{X}_{[k]})-f(\mathbf{Y}_{[k]})$ 
for $1\leq k\leq j$. 
For $k=2j-i+2$, 
\begin{equation*}
\begin{split}
&f(\mathbf{X}'_{[k]})-f(\mathbf{Y}'_{[k]})
\\
=&(f(\mathbf{X}_{[j]})+f(X_{j}Y_{j})+f(Y_{j}X_{i+1})
+f(\mathbf{X}_{[i+1,j+1]}))
\\
&-(f(\mathbf{Y}_{[j]})+f(Y_{j}X_{i+1})+f(\mathbf{X}_{[i+1,j]})
+f(X_{j}Y_{j})
\\
&+f(\mathbf{Y}_{[j,j+1]}))
\\
=&f(\mathbf{X}_{[j+1]})-f(\mathbf{Y}_{[j+1]}).
\end{split}
\end{equation*}
Therefore, $f(\mathbf{X}'_{[k]})-f(\mathbf{Y}'_{[k]})
=f(\mathbf{X}_{[k-(j-i+1)]})-f(\mathbf{Y}_{[k-(j-i+1)]})$ for $2j-i+2\leq k\leq n+j-i+3$. 
As a result, 
$\mathbf{F}(\mathbf{X})-\mathbf{F}(\mathbf{Y})$ is a subsequence of 
$\mathbf{F}(\mathbf{X}')-\mathbf{F}(\mathbf{Y}')$ which implies 
$\sigma(\mathbf{F}(\mathbf{X})-\mathbf{F}(\mathbf{Y}))\leq 
\sigma(\mathbf{F}(\mathbf{X}')-\mathbf{F}(\mathbf{Y}'))$. 

\textit{Case 3: $i>j$.} Similar to the discussion in Case 2, we select $\mathbf{Y}_{[j+1,i]}X_{i}$ 
as $\mathbf{z}$. 

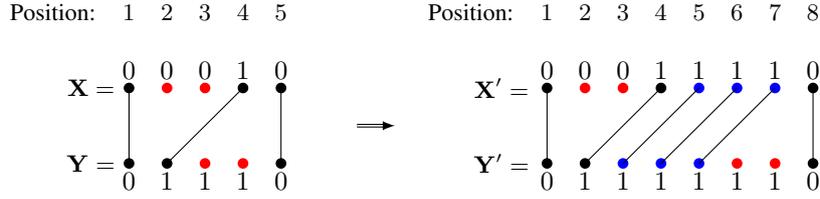
\begin{figure*}[htbp]
\centering
\begin{tikzpicture}

\node at (-0.5,1.5) {\small Position:};

\node at (0.5,1.5) {\small $1$};

\node at (1,1.5) {\small $2$};

\node at (1.5,1.5) {\small $3$};

\node at (2,1.5) {\small $4$};

\node at (2.5,1.5) {\small $5$};

\node at (0,0.5) {\small $\mathbf{X}=$};

\fill (0.5,0.5) circle (2pt);
\node[above] at (0.5,0.5) {$0$};

\fill[red] (1,0.5) circle (2pt);
\node[above] at (1,0.5) {$0$};

\fill[red] (1.5,0.5) circle (2pt);
\node[above] at (1.5,0.5) {$0$};

\fill (2,0.5) circle (2pt);
\node[above] at (2,0.5) {$1$};

\fill (2.5,0.5) circle (2pt);
\node[above] at (2.5,0.5) {$0$};

\node at (0,-0.5) {\small $\mathbf{Y}=$};

\fill (0.5,-0.5) circle (2pt);
\node[below] at (0.5,-0.5) {$0$};

\fill (1,-0.5) circle (2pt);
\node[below] at (1,-0.5) {$1$};

\fill[red] (1.5,-0.5) circle (2pt);
\node[below] at (1.5,-0.5) {$1$};

\fill[red] (2,-0.5) circle (2pt);
\node[below] at (2,-0.5) {$1$};

\fill (2.5,-0.5) circle (2pt);
\node[below] at (2.5,-0.5) {$0$};

\draw (0.5,0.5)--(0.5,-0.5);

\draw (2,0.5)--(1,-0.5);

\draw (2.5,0.5)--(2.5,-0.5);

\draw[-latex,double] (3.5,0)--(4,0);

\node at (5,1.5) {\small Position:};

\node at (6,1.5) {\small $1$};

\node at (6.5,1.5) {\small $2$};

\node at (7,1.5) {\small $3$};

\node at (7.5,1.5) {\small $4$};

\node at (8,1.5) {\small $5$};

\node at (8.5,1.5) {\small $6$};

\node at (9,1.5) {\small $7$};

\node at (9.5,1.5) {\small $8$};

\node at (5.4,0.5) {\small $\mathbf{X}'=$};

\fill (6,0.5) circle (2pt);
\node[above] at (6,0.5) {$0$};

\fill[red] (6.5,0.5) circle (2pt);
\node[above] at (6.5,0.5) {$0$};

\fill[red] (7,0.5) circle (2pt);
\node[above] at (7,0.5) {$0$};

\fill (7.5,0.5) circle (2pt);
\node[above] at (7.5,0.5) {$1$};

\fill[blue] (8,0.5) circle (2pt);
\node[above] at (8,0.5) {$1$};

\fill[blue] (8.5,0.5) circle (2pt);
\node[above] at (8.5,0.5) {$1$};

\fill[blue] (9,0.5) circle (2pt);
\node[above] at (9,0.5) {$1$};

\fill (9.5,0.5) circle (2pt);
\node[above] at (9.5,0.5) {$0$};

\node at (5.4,-0.5) {\small $\mathbf{Y}'=$};

\fill (6,-0.5) circle (2pt);
\node[below] at (6,-0.5) {$0$};

\fill (6.5,-0.5) circle (2pt);
\node[below] at (6.5,-0.5) {$1$};

\fill[blue] (7,-0.5) circle (2pt);
\node[below] at (7,-0.5) {$1$};

\fill[blue] (7.5,-0.5) circle (2pt);
\node[below] at (7.5,-0.5) {$1$};

\fill[blue] (8,-0.5) circle (2pt);
\node[below] at (8,-0.5) {$1$};

\fill[red] (8.5,-0.5) circle (2pt);
\node[below] at (8.5,-0.5) {$1$};

\fill[red] (9,-0.5) circle (2pt);
\node[below] at (9,-0.5) {$1$};

\fill (9.5,-0.5) circle (2pt);
\node[below] at (9.5,-0.5) {$0$};

\draw (6,0.5)--(6,-0.5);

\draw (7.5,0.5)--(6.5,-0.5);

\draw (8,0.5)--(7,-0.5);

\draw (8.5,0.5)--(7.5,-0.5);

\draw (9,0.5)--(8,-0.5);

\draw (9.5,0.5)--(9.5,-0.5);

\end{tikzpicture}
\caption{Apply segmentation technique to separate $l_{2}$ and $l_{3}$ for 
$\mathbf{X}=00010$ and $\mathbf{Y}=01110$.}
\label{Appendix2}
\end{figure*}

Fig. \ref{Appendix2} is an extreme example to illustrate its effectiveness. 
Let $\mathbf{X}=00010$ and $\mathbf{Y}=01110$. 
Then $\mathbf{X}_{[5]\backslash\{ l_{1},l_{2} \}}=\mathbf{Y}_{[5]\backslash\{ l_{3},l_{4} \}}$ 
where $l_{1}=2$, $l_{2}=3$, $l_{3}=3$, and $l_{4}=4$. At the beginning, 
$l_{1}$, $l_{2}$, $l_{3}$, and $l_{4}$ are in very close proximity to each other. 
We demonstrate the separation of $l_{2}$ and $l_{3}$. 
For $i=4$ and $j=2$, all matched $X_{\alpha}$ and $Y_{\beta}$ 
(i.e., $X_{1}$ with $Y_{1}$, $X_{4}$ with $Y_{2}$, and $X_{5}$ with $Y_{5}$) satisfy 
$\begin{cases}
\alpha\leq i
\\
\beta\leq j
\end{cases}$
or
$\begin{cases}
\alpha\geq i+1
\\
\beta\geq j+1
\end{cases}$. According to Case 3, we add $\mathbf{z}=\mathbf{Y}_{[3,4]}X_{4}=111$ between 
$X_{4}$ and $X_{5}$, as well as between $Y_{2}$ and $Y_{3}$, with 
initial separation of $l_{2}$ and $l_{3}$ completed. 
By implementing additional rounds of segmentation technique, we can obtain 
$\mathbf{U}$ and $\mathbf{V}$ that satisfy the requirements of Lemma \ref{M.l2.1}. 
}





\end{document}